\documentclass[preprint,12pt]{elsarticle}

\usepackage{amssymb,amsmath,amsthm,bm, todonotes}
\usepackage{ dsfont }
\usepackage{color}
\usepackage{hyperref}
\newcommand{\Prob} {\mbox{$\rm{Prob}$\,}}

\newcommand\E{\operatorname{E}}

\usepackage{lineno}

\journal{}

\begin{document}

\begin{frontmatter}

%% Title, authors and addresses

%% use the tnoteref command within \title for footnotes;
%% use the tnotetext command for theassociated footnote;
%% use the fnref command within \author or \address for footnotes;
%% use the fntext command for theassociated footnote;
%% use the corref command within \author for corresponding author footnotes;
%% use the cortext command for theassociated footnote;
%% use the ead command for the email address,
%% and the form \ead[url] for the home page:
%% \title{Title\tnoteref{label1}}
%% \tnotetext[label1]{}
%% \author{Name\corref{cor1}\fnref{label2}}
%% \ead{email address}
%% \ead[url]{home page}
%% \fntext[label2]{}
%% \cortext[cor1]{}
%% \address{Address\fnref{label3}}
%% \fntext[label3]{}

\title{Maximum likelihood estimators for scaled mutation rates in an equilibrium mutation-drift model}

%% use optional labels to link authors explicitly to addresses:
\author[label1]{Claus Vogl}
\ead{claus.vogl@vetmeduni.ac.at}
\author[label2]{Lynette C.\ Mikula}
\ead{lcm29@st-andrews.ac.uk}
\author[label3]{Conrad J.\ Burden}
\ead{conrad.burden@anu.edu.au}
\address[label1]{Department of Biomedical Sciences, Vetmeduni Vienna, Veterin\"arplatz 1, A-1210 Wien, Austria}
\address[label2]{Centre for Biological Diversity, School of Biology, University of St.\ Andrews, St Andrews KY16 9TH, UK}
\address[label3]{Mathematical Sciences Institute, Australian National University, Canberra, Australia}

\begin{abstract}
The stationary sampling distribution of a neutral decoupled Moran or Wright-Fisher diffusion with neutral mutations is known to first order for a general rate matrix with small but otherwise unconstrained mutation rates.  Using this distribution as a starting point we derive results for maximum likelihood estimates of scaled mutation rates from site frequency data under three model assumptions: a twelve-parameter general rate matrix, a nine-parameter reversible rate matrix, and a six-parameter strand-symmetric rate matrix.  The site frequency spectrum is assumed to be sampled from a fixed size population in equilibrium, and to consist of allele frequency data at a large number of unlinked sites evolving with a common mutation rate matrix without selective bias.  We correct an error in a previous treatment of the same problem \citet{BurdenTang17} affecting the estimators for the general and strand-symmetric rate matrices. The method is applied to a biological dataset consisting of a site frequency spectrum extracted from short autosomal introns in a sample of {\em Drosophila melanogaster} individuals.  
\end{abstract}

\begin{keyword}
mutation-drift model \sep decoupled Moran diffusion \sep Wright-Fisher diffusion \sep scaled mutation parameters \sep strand-symmetry 
\end{keyword}

\end{frontmatter}

%\linenumbers
%
%%%%%%%%%%%%%%%%%%%%%%%%%%%%%%%%%%%%%%%%%%%%%%%%%%%%%%%%%%%%
%
\section{Introduction}

A significant obstacle to inferring evolutionary parameters from allele frequency data is the paucity of accurate solutions to well-established stochastic population genetics models \citep{Tataru17}.  For instance, it should, in principle, be possible to infer scaled mutation rates from population allele frequencies observed in unlinked neutral genomic sites whose genealogies are independent due to the effects of recombination \citep{Rosenberg02}. As a minimal requirement, such inference requires solution of, say, a Wright-Fisher or Moran model with mutations, typically formulated in the diffusion limit. For the case of a stationary bi-allelic system, \citet{Vogl14} has demonstrated that this is feasible as the stationary distribution in the diffusion limit of the mutation-drift model is known to be a beta distribution and the corresponding sampling distribution is beta-binomial. This analysis has been extended to include a bi-allelic mutation-drift system with directional selection~\citep{VoglBergman15}.  

Inference of the complete $4 \times 4$ genomic mutation rate matrix from allele frequency data requires solution of a multi-allele mutation-drift model.  There is no known exact solution to the diffusion limit of the multi-allele mutation-drift model except for the case of parent-independent mutations, for which the stationary solution is known to be a Dirichlet distribution~\citep[][p394]{Wright69}, and the corresponding sampling distribution is Dirichlet-multinomial.  However, a recent extensive study of the moments of allele distributions under various model rate matrices \citep{Speed19} has demonstrated the shortcomings of the Dirichlet approximation in more general settings than the biologically unrealistic parent independent model.  

\textcolor{black}{
The scaled mutation rate in the diffusion limit, which is often termed $\theta$, is essentially the product of a per base substitution 
mutation rate per nucleotide site per generation ($u$) and an effective population size ($N$).  
In a study of silent states in protein-coding genes \citet[][Fig.~3b]{Lynch16} observe that $u$ decreases from about $\sim 10^{-8}$ to about $\sim 10^{-10}$ 
as $N$ ranges from $\sim 10^4$ to $\sim 10^8$ in eukaryotes.  This corresponds to scaled mutation rates which are $< 10^{-2}$.  
}
A promising approach therefore is to consider small-$\theta$ approximations either to solutions of the multi-allele mutation-drift model, or to the model itself.  

The stationary solution to the multi-allele mutation-drift diffusion with an arbitrary mutation rate matrix has been obtained to first order in $\theta$ \citep{BurdenTang16} by solving the forward Kolmogorov equation.  The corresponding sampling distribution has been determined by \citet{BurdenTang17} and verified by \citet{BurdenGriffiths19a} using a coalescent approach.  The identical sampling distribution has been derived independently by \citet{SchrempfHobolth17} from a boundary-mutation model~\citep{VoglBergman15} based on the decoupled Moran model.  

The purpose of the current paper is to explore the process of inferring a complete neutral mutation rate matrix from a spectrum of observed allele frequencies at independently evolving sites from the stationary sampling distribution. Our starting point is the small-$\theta$ sampling distribution described above. We obtain maximum likelihood (ML) estimates which can be efficiently computed under assumptions of (i) a general unconstrained rate matrix, (ii) a reversible rate matrix, and (iii) a strand-symmetric rate matrix.  In each case we construct combinations of rate matrix parameters, which have unbiased ML estimators, and determine which parameters necessarily have biased ML estimators. We also correct an error in \citet{BurdenTang17} in which the ML estimator of the rate matrix for a general unconstrained rate matrix is incorrectly stated.  

The format of the paper is as follows: In Section~\ref{sec:SamplingDistribution} the multi-allele mutation-drift diffusion is defined and the ${\mathcal O}(\theta)$ stationary sampling distribution is stated. The statement of the inference problem and a description of the form of the multi-allele frequency dataset for an effective haploid sample size of $M$ individuals at a total of $L$ sites (or loci) is described in Section~\ref{sec:siteFreqData}, together with a brief summary of how ML estimation will be implemented in subsequent sections.  Section~\ref{sec:QReparametrization} gives a reparametrization of the rate matrix in a form suitable for analysing reversible and non-reversible rate matrices. Derivation of ML estimates for the case of a general unconstrained rate matrix and for the case of a rate matrix constrained to be reversible are given in Section~\ref{sec:MaxLikelihoodEstimates}. The strand-symmetric case is covered in Section~\ref{sec:strandSymmetry}. The theory is applied to a dataset extracted from short autosomal introns of 197 {\em Drosophila melanogaster} individuals at 218,942 genomic sites in Section~\ref{sec:Application}\textcolor{black}{, and simulations exploring the accuracy and applicability of the small-$\theta$ approximation can be found in Section~\ref{sec:Simulation}}. Conclusions are drawn in Section~\ref{sec:Conclusions}, while an Appendix is devoted to technical details of numerical optimisation.  
%
%%%%%%%%%%%%%%%%%%%%%%%%%%%%%%%%%%%%%%%%%%%%%%%%%%%%%%%%%%%%
%
\section{Stationary sampling distribution for a general rate matrix}
\label{sec:SamplingDistribution}

\textcolor{black}{
We will begin our derivations by considering a $K$-allele neutral decoupled Moran model~\citep{BaakeBialowons08,EtheridgeGriffiths09,VoglClemente12} or haploid Wright-Fisher model~\citep[][p55]{Etheridge12} with scaled rate matrix 
\begin{equation}	\label{continuumScaling}
Q_{ij} = \begin{cases}
N(u_{ij} - \delta_{ij}) & \text{(decoupled Moran)}\\
2N(u_{ij} - \delta_{ij}) & \text{(Wright-Fisher)},
\end{cases}
\end{equation}
where $N$ is the haploid (effective) population size and $u_{ij}$ are the per generation mutation rates.
The diffusion limit of this process is defined by the simultaneous limits $N \rightarrow \infty$ and $u_{ij} \rightarrow \delta_{ij}$ for fixed $Q_{ij}$ and can be described by the backward generator 
\begin{equation}		\label{backwardGenerator}
\frac{1}{2} \sum_{i, j = 1}^K x_i(\delta_{ij} - x_j) \frac{\partial^2}{\partial x_i \partial x_j} + \frac{1}{2} \sum_{i = 1}^K Q_{ij} x_j \frac{\partial}{\partial x_i}. 
\end{equation}
where $x_{i}$ and $x_{j}$ denote the respective allele frequencies.
}

In its most general form the $K \times K$ rate matrix $\mathbf{Q}$ is constrained by 
\begin{equation}    \label{QProperties}
Q_{ij} \ge 0, \quad \text{for }i \ne j; \qquad \sum_{j = 1}^K Q_{ij} =0, 
\end{equation} 
implying that $K(K - 1)$ parameters are required to specify $\mathbf{Q}$.  
Let us assume that $\mathbf{Q}$ has a unique stationary state $\pi^{\rm T} = (\pi_1 \ldots \pi_K)$ satisfying 
\begin{equation}    \label{PiProperties}
\pi_i \ge 0, \quad \sum_{i = 1}^K \pi_i = 1, \quad \sum_{i = 1}^K \pi_i Q_{ij} = 0.    
\end{equation}
A sufficient condition for a unique $\pi^{\rm T}$ to exist is \textcolor{black}{$Q_{ij} > 0$ for all $i \ne j$}. One would expect this to include any biologically realistic model.  

Suppose we further assume small scaled mutation rates, that is, assume the off-diagonal elements of $\mathbf{Q}$ to be ${\cal O}(\theta)$ for some small parameter $\theta$. The sampling distribution for a finite sample of $M$ individuals randomly and independently drawn from the population \textcolor{black}{of size $N$} has been obtained to first order in $\theta$ by 
\citet[][Eq.~(35)]{BurdenTang17} from an approximate solution to the forward Kolmogorov equation corresponding to the generator in Eq.~(\ref{backwardGenerator}) and by \citet[][Theorem~1]{BurdenGriffiths19a} from the coalescent. An identical distribution has also been given by \citet{SchrempfHobolth17} using the boundary-mutation model as a starting point. Let 
\begin{equation} \label{alleleOccupancy}
\mathbf{Y} = (Y_1, \ldots, Y_K), \qquad \sum_{i = 1}^K Y_i = M, 
\end{equation}
be the occupancy of alleles in a population sample of size $M$, assuming stationarity. Then the stationary sampling distribution is, to first order in $\theta$, 
\begin{equation}\label{eq:general_stationary_distribution}
\Pr(\mathbf{Y} = \mathbf{y} \mid M, \mathbf{Q}) = 
  \begin{cases} 
  \displaystyle \pi_i\left(1 - H_{M-1} \sum_{j\neq i} Q_{ij}\right)  + \mathcal{O}(\theta^{2}), & y_i=M,y_{j\neq i}=0; \\ \\
  \displaystyle{\frac{\pi_i Q_{ij}}{y_j} + \frac{\pi_j Q_{ji}}{y_i}} + \mathcal{O}(\theta^{2}), & 1\leq y_i,y_{j\neq i}\leq M-1 \\
 												 &\text{  and } y_i+y_j=M\,; \\ \\
 \mathcal{O}(\theta^{2}), & y_i, y_j, y_k >0 \text{ for distinct $i$, $j$, $k$,} 
\end{cases}
\end{equation}
where
\begin{equation} 
H_{M - 1} = \sum_{y = 1}^{M- 1} \frac{1}{y}.
\end{equation}
This distribution is a generalisation of special cases corresponding to situations where the stationary distribution of the neutral Moran or Wright-Fisher diffusion is known exactly. The corresponding 2-allele case is quoted in \citet[][Eq.~(29)]{Vogl14}, and the case of multi-allelic parent-independent rate matrix is given in \citet[][Eq.~(10)]{RoyChoudhuryWakeley10}.  Both of these special cases correspond to reversible rate matrices, for which $\pi_i Q_{ij} = \pi_j Q_{ji}$, leading to simplification of the second line in Eq.~(\ref{eq:general_stationary_distribution}).  
%
%%%%%%%%%%%%%%%%%%%%%%%%%%%%%%%%%%%%%%%%%%%%%%%%%%%%%%%%%%%%
%
\section{Site frequency data}
\label{sec:siteFreqData}

Our aim is to estimate the scaled mutation rate matrix $\mathbf{Q}$ from 
a dataset in the form of a site frequency spectrum obtained by sampling $L$ independent neutrally evolving loci 
within multiple alignments of $M$ genomes. The obvious application is to the genomic alphabet $\{A, T, G, C\}$ of $K = 4$ letters, with the loci being neutral genomic sites such as fourfold degenerate sites within codons or short intron sites~\citep{VoglBergman15}. Such a dataset can be achieved in principle from a sample of $M/2$ diploid, monoecious individuals, with the sites chosen to be sufficiently separated so as to have independent coalescent trees due to recombination. In terms of the allele occupancy counts defined by Eq.~(\ref{alleleOccupancy}), set  
\begin{equation}   \label{eq:L_definitions1}
\begin{split}
L_i & =  \text{the number of sites with $Y_i = M$, $Y_{j \ne i} = 0$};  \\
L_{ij}(y) & =  \text{the number of sites with $Y_i = y$, $Y_{j \ne i} = M - y$, $Y_{k \ne i, j} = 0$}; \\
\end{split}
\end{equation}
for $1 \le i, j, k \le K$ and $y = 1, \ldots M - 1$. Also define 
\begin{equation}    \label{eq:L_definitions2}
\begin{split}
L_{ij} &= \sum_{y = 1}^{M - 1} L_{ij}(y);  \\
L_{\rm P} &=  \sum_{1 \le i < j \le K} L_{ij}.
\end{split}
\end{equation}
Thus $L_i$ counts the number of non-segregating sites with allele $i$, $L_{i j}$ counts the number of bi-allelic sites with alleles $i$ and $j$, and $L_{\rm P}$ counts the total number of bi-allelic sites. Since Eq.~(\ref{eq:general_stationary_distribution}) implies that tri-allelic, tetra-allelic, etc.\ sites only occur with probability ${\mathcal O}(\theta^2)$, we further assume that all sites are either non-segregating (with probability ${\mathcal O}(1)$) or bi-allelic (with probability ${\mathcal O}(\theta)$), and hence \textcolor{black}{the total number of sites is} 
\begin{equation}
L = \sum_i L_i +L_{\rm P}.  
\end{equation}
Note that under the model defined by the distribution Eq.~(\ref{eq:general_stationary_distribution}), $L_i$ and $L_{ij}(y)$ are random variables, whereas $L$ is set by  experimental design and is not a random variable.  

\textcolor{black}{
Let the vector of random site counts
\begin{equation}	\label{LAsAVector}
\mathbf{L} = (L_i, L_{ij}(y)), \quad 1 \le i \le K; \, 1 \le j < j \le K; \, y = 1, \ldots, M - 1,  
\end{equation}
listed in Eq.~(\ref{eq:L_definitions1}) take observed values $\boldsymbol\ell = (l_i, l_{ij}(y))$.  
To first order in $\theta$, we obtain from Eq.~(\ref{eq:general_stationary_distribution}) the likelihood function 
\begin{eqnarray} \label{eq:likelihood}
{\cal L}(\mathbf{Q} \mid \boldsymbol\ell) & := & \Pr\left(\mathbf{L} = \bf{\boldsymbol\ell} \mid \mathbf{Q} \right) \nonumber \\
& = & \frac{L!}{(\prod_i l_i!) (\prod_{i < j} \prod_{y = 1}^{M - 1} l_{ij}(y)!)} \nonumber \\
& & \quad \times 	\prod_{i = 1}^K \left\{\pi_i\left(1 - H_{M-1} \sum_{j\neq i} Q_{ij}\right)\right\}^{l_i} \nonumber \\
& & \quad \times	\prod_{1 \le i < j \le K} \prod_{y = 1}^{M - 1} \left\{{\frac{\pi_i Q_{ij}}{M - y} + \frac{\pi_j Q_{ji}}{y}}\right\}^{l_{ij}(y)}.  
\end{eqnarray}
Note that this is a multinomial distribution in $\mathbf{L}$. 
}

The general form of a multinomial distribution is 
\begin{equation}  \label{generalMultinomial}
\Prob(\mathbf{L} = \boldsymbol\ell) = \frac{ \textcolor{black}{L!} }{\prod_\alpha l_\alpha!} \prod_\alpha p_\alpha^{l_\alpha} = {L \choose \boldsymbol\ell} \prod_\alpha p_\alpha^{l_\alpha}, 
\end{equation}
where $\sum_{\alpha = 1}^n l_\alpha = L$ for a fixed number of categories $n$ and parameters $p_\alpha$, with $\alpha = 1, \ldots, n$ constrained by 
\footnote{In Eq.~(\ref{eq:likelihood}) the number of categories is $n = K + \tfrac{1}{2}K(K - 1)(M - 1)$.  Later in this paper the general properties of multinomials quoted here will be applied to marginal distributions, which are multinomials with lesser values of $n$.} 
\begin{equation} \label{eq:pConstraints}
p_\alpha > 0, \qquad \sum_{\alpha = 1}^n  p_\alpha = 1.  
\end{equation}
In general, $E(L_\alpha) = L p_\alpha$, so the random variables 
\begin{equation}	\label{unbiasedPHat}
\hat{p}_\alpha = \frac{L_\alpha}{L}, 
\end{equation}
are unbiased estimators of $p_\alpha$.  By writing Eq.~(\ref{generalMultinomial}) in the canonical form 
\begin{equation}		\label{exponentialFam}
\Prob(\mathbf{L} = \boldsymbol\ell) = {L \choose \boldsymbol\ell} p_n^L \exp\left(\sum_{\beta = 1}^{n - 1} l_\beta \log\frac{p_\beta}{p_n} \right), 
\end{equation} 
one sees that the multinomial with fixed $L$ and unknown $p_\alpha$ constitutes an exponential family of distributions with sufficient statistics $L_1, \ldots, L_{n - 1}$ (sufficiency can easily be seen via the Neyman factorisation theorem ~\citep[][pp~318, 341]{HoggCraig95}).  \textcolor{black}{Essentially this means that all the information from the data needed to construct any estimator of $p_\alpha$ is known through $L_1, \ldots, L_{n - 1}$; the probability of the data given $L_1, \ldots, L_{n - 1}$ is therefore independent of $p_\alpha$ itself.}

Furthermore, if Eq.~(\ref{eq:pConstraints}) are the {\em only} constraints on the $p_\alpha$, then $L_1, \ldots, L_{n - 1}$ are a minimal complete set of sufficient statistics (completeness via Definition 3.19 of \citet[][p~50]{Keener2010theoretical}) \textcolor{black}{and the natural parameter space maps identifiably onto that of the estimators.} The theory of exponential families tells us that \textcolor{black}{$\hat{p_\alpha}$} defined by Eq.~(\ref{unbiasedPHat}) are in fact uniformly minimum variance unbiased estimators. 
\textcolor{black}{
This can be shown by application of the Rao-Blackwell theorem as shown in Theorem 4.4 of \citet[][p~62]{Keener2010theoretical}. The argument via completeness in the referenced proof is often given separately as the Lehmann-Sheffe theorem, which can also be verified with a straightforward calculation using Lagrange multipliers. This shows that these estimators are also unique ML estimators.} We will refer to such multinomials as being {\em flat}.   

However, the analogous parameters in Eq.~(\ref{eq:likelihood}) are functions of elements of the rate matrix $\mathbf{Q}$, which are themselves subject to nontrivial constraints via Eqs.~(\ref{QProperties}) and (\ref{PiProperties}). In this case, the number of independent parameters is less than the number of sufficient statistics and the exponential family is said to be {\em curved} \citep[][Chapter~5]{Keener2010theoretical}. Many of the standard results pertaining to {\em flat} exponential families do not generalise to {\em curved} exponential families. Importantly for our case, the estimators defined by Eq.~(\ref{unbiasedPHat}) are still unbiased but, in general, are {\em not} ML estimators of $p_\alpha$.  

In the following sections we explore the problem of determining ML estimators under various model restrictions on the rate matrix $\mathbf{Q}$. In most cases exact analytic formulae for ML estimates of the complete set of parameters are intractable. However, by judicious use of reparametrization we are able exploit the Neyman factorisation theorem~\citep[][pp~318, 341]{HoggCraig95} to factor Eq.~(\ref{eq:likelihood}) into (i) flat marginal multinomials, from which minimum variance unbiased ML estimators can be obtained for certain combinations of parameters, and (ii) a quotient depending on the remaining parameters, for which ML estimators can be determined numerically.  
We begin by introducing a reparametrization of the rate matrix into reversible and non-reversible parts, which will enable us to specify a convenient minimal set of independent parameters of the general rate matrix $\mathbf{Q}$.  
%
%%%%%%%%%%%%%%%%%%%%%%%%%%%%%%%%%%%%%%%%%%%%%%%%%%%%%%%%%%%%
%
\section{Reparametrization of the rate matrix}
\label{sec:QReparametrization}

\textcolor{black}{
Recall the defining properties of the rate matrix $\mathbf{Q}$ and its stationary distribution $\mathbf{\pi^T}$, Eqs.~(\ref{QProperties}) and (\ref{PiProperties}).}

Define the parameters 
\begin{equation}	\label{CPhiDefns}
C_{ij} = \tfrac{1}{2} (\pi_i Q_{ij} + \pi_j Q_{ji}), \qquad \Phi_{ij} = \tfrac{1}{2} (\pi_i Q_{ij} - \pi_j Q_{ji}), 
\end{equation}
for $i, j = 1, \ldots, K$.  One easily checks that 
\begin{equation}	\label{QinTermsOfCPhi}
Q_{ij} =  (C_{ij} + \Phi_{ij})/\pi_i,   
\end{equation}
where the first term is the reversible part of the rate matrix, $Q_{ij}^{\rm GTR} = C_{ij}/\pi_i$, and the second term, $Q_{ij}^{\rm flux} = \Phi_{ij}/\pi_i$ represents a 
flux $\Phi_{ij}$ of probability per unit time from allele $i$ to allele $j$~\citep{BurdenTang17}.  

\textcolor{black}{It follows} that $C_{ij}$ are the 
elements of a symmetric $K \times K$ matrix satisfying $\sum_{j = 1}^K C_{ij} = 0$, and that $\Phi_{ij}$ are the elements of an antisymmetric 
$K \times K$ matrix satisfying $\sum_{j = 1}^K \Phi_{ij} = 0$.  This last constraint is a statement that fluxes of probability out of any allele 
to the remaining $K - 1$ alleles must be zero.  

% Figure fig:ProbFluxes
\begin{figure}[t!]
\begin{center}
\centerline{\includegraphics[width=0.48\textwidth]{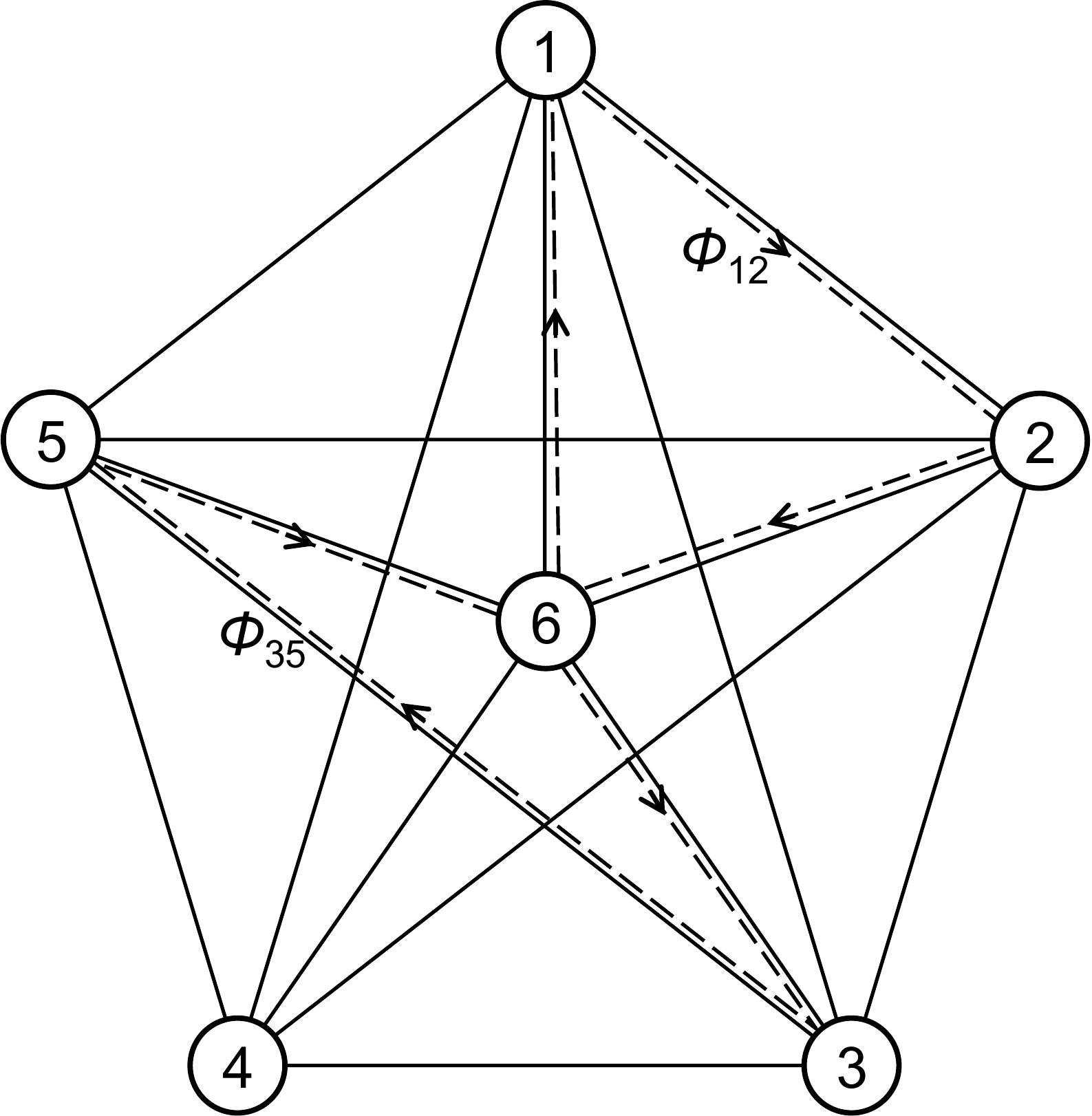}}
\caption{Probability fluxes through the edges of the simplex spanned by $K = 6$ alleles. The independent $\Phi_{ij}$ for all $1 \le i < j \le K - 1$ contribute  probability fluxes around paths $K \rightarrow i \rightarrow j \rightarrow K$ as shown for $\Phi_{12}$ and $\Phi_{35}$.  The remaining radial fluxes $\Phi_{iK}$ for $i \ne K$ are obtained by summing fluxes from all such triangular paths passing through $i$ and $K$.} 
\label{fig:ProbFluxes}
\end{center}
\end{figure}

Clearly the stationary distribution $\pi_i$ contains $K - 1$ independent degrees of freedom, and the parameters $C_{ij}$ contribute $\frac{1}{2}K(K - 1)$ independent degrees of freedom. To understand the number of independent elements of $\Phi_{ij}$, consider Fig.~\ref{fig:ProbFluxes} illustrating the case $K = 6$.  
Placing the $K$-th allele at the centre of the diagram, we observe that the $\frac{1}{2}(K - 1)(K - 2)$ closed paths $K \rightarrow i \rightarrow j \rightarrow K$ for $1 \le i < j \le K - 1$ define independent fluxes $\Phi_{ij}$, and that the flux along any radial edge $i \rightarrow K$ for $1 \le i \le K - 1$ can be obtained by conserving flux at the vertex $i$.  To summarise, the general rate matrix $\mathbf{Q}$ can be parameterised using the following minimal set of parameters: 
\begin{equation}		\label{parameterList}
\begin{array}{lrr}
\pi_i, & i = 1, \ldots, K - 1: & K - 1\mbox{ parameters}; \\
C_{ij}, & 1 \le i < j \le K: & \tfrac{1}{2}K(K - 1) \mbox{ parameters}; \\
\Phi_{ij}, & 1 \le i < j \le K - 1: & \tfrac{1}{2}(K - 1)(K - 2) \mbox{ parameters},
\end{array}
\end{equation}
with the remaining, dependent parameters given by
\begin{equation}	\label{dependentParams}
\begin{split}
\pi_K &= \sum_{i = 1}^{K - 1} \pi_i, \\
C_{ji} &= C_{ij}, \quad 1 \le i < j\le K, \\
C_{ii} &= \sum_{j \ne i} C_{ij} \quad 1 \le i \le K, \\
\Phi_{ji} & = -\Phi_{ij}, \quad 1 \le i < j < K, \\
\Phi _{iK} &= -\Phi_{Ki} = \sum_{j = 1}^{i - 1} \Phi _{ji} - \sum_{j = i + 1}^{K - 1} \Phi _{ij},  \quad 1 \le i \le K - 1.
\end{split}
\end{equation}
The total number of independent parameters is $K(K - 1)$, as required.  

%
%%%%%%%%%%%%%%%%%%%%%%%%%%%%%%%%%%%%%%%%%%%%%%%%%%%%%%%%%%%%
%
\section{Maximum likelihood estimates}
\label{sec:MaxLikelihoodEstimates}

%
%%%%%%%%%%%%%%%%%%%%%%%%%%%%%%%%%%%%%%%%%%%%%%%%%%%%%%%%%%%%
%
\textcolor{black}{
We can now use the theory from Section~\ref{sec:siteFreqData} to discuss the properties of ML estimators of the (reparametrized) rate matrix in three biologically relevant scenarios. 
}

\subsection{$K = 2$ alleles}
\label{sec:MaxLikelihoodEstimatesKEquals2}

Because the $2 \times 2$ rate matrix for the bi-allelic mutation-drift model is necessarily reversible, there are no probability fluxes and therefore only two independent parameters, $\pi_1$ and $C_{12}$.  The ML estimators of these parameters were derived by \citet{Vogl14}. Here we re-derive the estimators in a way which readily generalises to both the general $K$-allele model and the reversible $K$-allele model.  

From Eqs.~(\ref{eq:likelihood}), (\ref{QinTermsOfCPhi}) and (\ref{dependentParams}) we have 
\begin{eqnarray} \label{eq:probKEquals2}
\lefteqn{\Pr\left(L_1 = l_1,  L_2 = l_2, L_{12}(y) = l_{12}(y) \mid \pi_1, C_{12} \right)} \qquad\qquad\qquad\qquad  \nonumber \\
& = & g(l_1, l_2, l_{12} ; \pi_1, C_{12}) \times h(\vec{l}_{12}), 
\end{eqnarray}
where 
\begin{eqnarray}	\label{gKEquals2}
\lefteqn{g(l_1, l_2, l_{12} ; \pi_1, C_{12}) }   \nonumber \\
 & = & \frac{L!}{l_1! l_2! l_{12}!} (\pi_1 - H_{M - 1} C_{12})^{l_1}  (1 - \pi_1 - H_{M - 1} C_{12})^{l_2} (2 H_{M - 1}C_{12})^{l_{12}}, 
\end{eqnarray}
\begin{equation}
h(\vec{l}_{12}) = \frac{1}{(2 H_{M - 1})^{l_{12}}} \frac{l_{12}!}{\prod_{y = 1}^{M - 1} l_{12}(y)!} \prod_{y = 1}^{M - 1} \left( \frac{1}{M - y} + \frac{1}{y} \right) ^{l_{12}(y)},  
\end{equation}
and 
\begin{equation}
\vec{l}_{12} = (l_{12}(1), \ldots, l_{12}(M - 1)). 
\end{equation}
Since $h(\vec{l}_{12})$ does not depend on $\pi_1$ or $C_{12}$, the Neyman factorisation theorem for multiple parameters~\citep[][pp~341]{HoggCraig95} necessitates that $L_1, L_2$ and $L_{12}$ are sufficient statistics for estimating $\pi_1$ and $C_{12}$.  Then, since 
\begin{equation}	\label{sumOverYGives1}
\sum_{\{\vec{l}_{12} : \,\sum_y l_{12}(y) = l_{12}\}} h(\vec{l}_{12}) = \left\{ \frac{1}{2 H_{M - 1}} \sum_{y = 1}^{M - 1}\left( \frac{1}{M - y} + \frac{1}{y} \right) \right\}^{l_{12}} = 1, 
\end{equation}
the marginal probability in $L_1, L_2$ and $L_{12}$ is simply 
\begin{equation}
\Pr\left(L_1 = l_1,  L_2 = l_2, L_{12} = l_{12} \mid \pi_1, C_{12} \right) = g(l_1, l_2, l_{12} ; \pi_1, C_{12}).  
\end{equation}  
Eq.~(\ref{gKEquals2}) is a flat family of trinomial distributions with two independent parameters.  It follows from Eq.~(\ref{unbiasedPHat}) and the discussion following Eq.~(\ref{exponentialFam}) that $\hat{p}_1 = L_1/L$ and $\hat{p}_{12} = L_{12}/L$ are minimum variance, unbiased, 
ML estimators of  $p_1 := \pi_1 - H_{M - 1} C_{12}$ and $p_{12} := 2 H_{M - 1}C_{12}$ respectively.  The required ML estimators are then 
\begin{equation}
\hat{C}_{12} = \frac{L_{12}}{2LH_{M - 1}}, \quad  \hat{\pi}_1= \frac{L_1 + \frac{1}{2}L_{12}}{L},
\end{equation}
agreeing with \citet[][Eqs.~(36) and (37)]{Vogl14}.  By linearity they are also unbiased.
%
%%%%%%%%%%%%%%%%%%%%%%%%%%%%%%%%%%%%%%%%%%%%%%%%%%%%%%%%%%%%
%
\subsection{General $K$-allele model}
\label{sec:MaxLikelihoodEstimatesKGeneral}

Returning to the $K$-allele mutation-drift model for a general $K \times K$ rate matrix, we have, from Eqs.~(\ref{eq:likelihood}), (\ref{QinTermsOfCPhi}) 
and the properties of $C_{ij}$ and $\Phi_{ij}$, 
\begin{eqnarray} \label{eq:probKGeneral}
\lefteqn{\Pr\left(\mathbf{L} = \boldsymbol\ell \mid \pi_i, C_{ij}, \Phi_{ij} \right)} \nonumber \\
& = & \frac{L!}{(\prod_i l_i!) (\prod_{i < j} \prod_{y = 1}^{M - 1} l_{ij}(y)!)} 
	 	\prod_{i = 1}^K \left(\pi_i - H_{M-1} \sum_{j\neq i} C_{ij}\right)^{l_i} \nonumber \\
& & \quad \times	\prod_{1 \le i < j \le K} \prod_{y = 1}^{M - 1} 
					\left\{C_{ij}\left({\frac{1}{M - y} + \frac{1}{y}}\right) + \Phi_{ij}\left({\frac{1}{M - y} - \frac{1}{y}}\right)\right\}^{l_{ij}(y)},  
\end{eqnarray}
where the vector $\mathbf{L}$ of random variables represents the complete set of counts in Eq.~(\ref{LAsAVector}).  

Our aim is to choose a parametrization that will enable us to exploit the Neyman factorisation theorem. To this end we define 
\begin{equation}	\label{c_phiDefns}
\begin{split}
C &= \sum_{1 \le i < j \le K} C_{ij}; \\
\pi_i' &= \frac{\pi_i - H_{M - 1} \sum_{j \ne i}C_{ij}}{1 - 2H_{M - 1}C}, \quad i = 1, \ldots, K; \\
c_{ij} &= \frac{C_{ij}}{C}, \quad i, j = 1, \ldots, K;  \\
\phi_{ij} &= \frac{\Phi_{ij}}{C}, \quad i, j = 1, \ldots, K.  
\end{split}
\end{equation}  
One easily checks that $\sum_{i = 1}^K \pi_i' = 1$, and therefore $K - 1$ of $\pi_i'$ are independent; that $\sum_{i<j} c_{ij} = 1$, and therefore $\frac{1}{2}K(K - 1) - 1$ of the $c_{ij}$ are independent; and that, by analogy with the $\Phi_{ij}$, there are $\frac{1}{2}(K - 1)(K - 2)$ independent rescaled fluxes $\phi_{ij}$.  Together with $C$ this gives a total of $K(K - 1)$ independent parameters, as required. In the following we choose for the set of independent parameters
\begin{equation}	\label{independentParams}
\{\pi_1', \ldots, \pi_{K - 1}', C\} \cup \left(\{c_{ij}: 1 \le i < j \le K\}\backslash\{c_{K-1,K}\}\right) \cup \{\phi_{ij}: 1 \le i < j \le {K - 1} \}.  
\end{equation}
The remaining, dependent, parameters in Eq.~(\ref{c_phiDefns}) are then defined as  
\begin{equation} \label{cKKminus1AndPhi_ik}
\begin{split}
c_{K - 1,K} &= 1 - \sum_{1 \le i < j \le K - 1} c_{ij} - \sum_{i = 1}^{K - 2} c_{iK}, \\
\phi _{iK} &= \sum_{j = 1}^{i - 1} \phi _{ji} - \sum_{j = i + 1}^{K - 1} \phi _{ij},  \quad 1 \le i \le K - 1.
\end{split}
\end{equation}

\textcolor{black}{
Let us further define the vector of bi-allelic counts as 
\begin{equation}
\boldsymbol\ell_{\rm P} = (l_{ij}(y)), \qquad 1 \le i < j \le K, \, y = 1, \ldots, M - 1, 
\end{equation} 
and the sum
\begin{equation}
l_{\rm P}=\sum_{i < j} \sum_y l_{ij}(y)
\end{equation}
}

This reparametrization gives Eq.~(\ref{eq:probKGeneral}) as  
\begin{eqnarray} \label{eq:probKGeneralReparam}
\lefteqn{\Pr\left(\mathbf{L} = \boldsymbol\ell \mid \pi_i', C, c_{ij}, \phi_{ij} \right)}  \qquad\qquad  \nonumber \\
& = & g(l_1, \ldots, l_K, l_{\rm P} ; \pi_1', \ldots, \pi_{K - 1}', C) \times h(\boldsymbol\ell_{\rm P}; c_{ij}, \phi_{ij}), 
\end{eqnarray}
where
\begin{eqnarray}
\lefteqn{g(l_1, \ldots, l_K, l_{\rm P} ; \pi_1', \ldots, \pi_{K - 1}', C) }   \nonumber \\
 & = & \frac{L!}{(\prod_{i = 1}^K \textcolor{black}{l_i!})l_{\rm P}!}  \left(\prod_{i = 1}^{K - 1}\left\{(1 - 2H_{M - 1} C)\pi_i'\right\}^{l_i}  \right) \nonumber \\
 & & 				\qquad \times		\left\{(1 - 2H_{M - 1} C)\left(1 - \sum_{i = 1}^{K - 1} \pi_i' \right)\right\}^{l_K} (2 H_{M - 1}C)^{l_{\rm P}}, 
\end{eqnarray}
and 
\begin{eqnarray}	\label{hGeneralReparam}
\lefteqn{h(\boldsymbol\ell_{\rm P}; c_{ij}, \phi_{ij}) = \frac{1}{(2 H_{M - 1})^{l_{\rm P}}} \frac{l_{\rm P}!}{\prod_{i < j} \prod_{y = 1}^{M - 1} l_{ij}(y)!}} \nonumber \\
	& & \qquad\times  \prod_{1 \le i < j \le K} \prod_{y = 1}^{M - 1} 
					\left\{ c_{ij} \left( \frac{1}{M - y} + \frac{1}{y} \right) + \phi_{ij}\left( \frac{1}{M - y} - \frac{1}{y} \right) \right\}^{l_{ij}(y)},   \nonumber \\
\end{eqnarray}
where we have used the notational convention of Eq.~(\ref{cKKminus1AndPhi_ik}).   

Since $h(\boldsymbol\ell_{\rm P}; c_{ij}, \phi_{ij})$ is independent of $C$ and $\pi_i'$, the Neyman factorisation theorem necessitates that $\{L_1, \ldots, L_K, L_{\rm P}\}$ is a sufficient set statistics for jointly estimating $C$ and $\pi_1', \ldots, \pi_{K - 1}'$. Following the same line of argument as for the $K = 2$ case, since 
\begin{eqnarray}
\lefteqn{\sum_{\{\boldsymbol\ell_{\rm P} : \,\sum_{i < j} \sum_y l_{ij}(y) = l_{\rm P}\}} h(\boldsymbol\ell_{\rm P}; c_{ij}, \phi_{ij})} \nonumber \\
& = &\left\{ \frac{1}{2 H_{M - 1}} 
			\sum_{1 \le i < j \le K} \sum_{y = 1}^{M - 1} 
			\left[ c_{ij} \left( \frac{1}{M - y} + \frac{1}{y} \right) + \phi_{ij}\left( \frac{1}{M - y} - \frac{1}{y} \right) \right] \right\}^{l_{\rm P}} \nonumber \\
& = & 1, 
\end{eqnarray}
the marginal probability in $\{L_1, \ldots, L_K, L_{\rm P}\}$ is simply 
\begin{equation}
\Pr\left(\textcolor{black}{L_1 = l_1,...,L_K=l_K}, L_{\rm P} = l_{\rm P} \mid \pi_i', C \right) = g(l_1, \ldots, l_K, l_{\rm P} ; \pi_1', \ldots, \pi_{K - 1}', C).  
\end{equation}  
This is a flat family of multinomial distributions with $K + 1$ categories and $K$ independent parameters. It follows from Eq.~(\ref{unbiasedPHat}) and the discussion following Eq.~(\ref{exponentialFam}) that $\hat{p}_1 = L_1/L$ to $\hat{p}_{K - 1} = L_{K - 1}/L$ are minimum variance, unbiased, ML estimators of $p_1 := (1 - 2H_{M - 1} C)\pi_1'$ to $p_{K - 1} := (1 - 2H_{M - 1} C)\pi_{K - 1}'$ respectively, and that $\hat{p}_{\rm P} = L_{\rm P}/L$ is a minimum variance, unbiased, ML estimator of $p_{\rm P} := 2 H_{M - 1}C$.  Thus we obtain the ML estimators 
\begin{equation}	\label{C_piPrime_Estimates}
\hat{C} = \frac{L_{\rm P}}{2LH_{M - 1}}, \qquad \hat{\pi}_i' = \frac{L_i}{L - L_{\rm P}} = \frac{L_i}{\sum_{j = 1}^K L_j}, \quad i = 1, \ldots, K. 
\end{equation}
Note that $\hat{C}$ is unbiased, but that the $\hat{\pi}_i' $ are biased.

\textcolor{black}{
Incidentally, the number of observed non-segregating sites $L - L_{\rm P}$ in the denominator of $\hat{\pi}_i'$ 
is highly unlikely to be zero by the following argument.  If the elements of $\mathbf Q$ are ${\cal O}(\theta)$, 
then from Eq.~(\ref{eq:likelihood}) $\Prob(L - L_{\rm P} = 0) = {\cal O}((\theta H_{M - 1})^L) = {\cal O}((\theta \log M)^L)$, 
which, for large $L$, is infinitesimal provided $\log(M)<1/\theta$.  With a conservative upper bound $\theta < 0.1$, 
this corresponds to a generous upper limit to the population sample of $M < 2.2 \times 10^4$.  Conversely, 
if all sites are observed to be segregating, then the small-$\theta$ approximation is unlikely to be appropriate.  
}

As it stands there is no practical way to factorise $h(\boldsymbol\ell_{\rm P}; c_{ij}, \phi_{ij})$ further into distinct subsets of the factors occurring in Eq.~(\ref{hGeneralReparam}) depending on corresponding distinct subsets of parameters because of the interdependencies in Eq.~(\ref{cKKminus1AndPhi_ik}). In practice, the ML estimate of the full rate matrix is completed by numerically maximising $h(\boldsymbol\ell_{\rm P}; c_{ij}, \phi_{ij})$ over its $(K - 1)^2 - 1$ independent parameters, and reconstructing $\mathbf{Q}$ via Eqs.~(\ref{QinTermsOfCPhi}), (\ref{c_phiDefns}) and (\ref{C_piPrime_Estimates}).  

%
%%%%%%%%%%%%%%%%%%%%%%%%%%%%%%%%%%%%%%%%%%%%%%%%%%%%%%%%%%%%
%
\subsection{General time reversible $K$-allele model}
\label{sec:MaxLikelihoodEstimatesKGTR}

The general time reversible rate matrix is defined to be the general rate matrix with the further constraint that $\pi_i Q_{ij} = \pi_j Q_{ji}$ for all 
$i, j = 1, \ldots, K$.  This is equivalent to {\em a priori} setting all $\Phi_{ij} \equiv 0$ in the parametrization of Eq.~(\ref{parameterList}).  

With 
this simplification the $C_{ij}$ decouple, and \textcolor{black}{further using
\begin{equation}	\label{vecLDefn}
\vec{l}_{ij} = (l_{ij}(1), \ldots, l_{ij}(M - 1)),  
\end{equation}
}
Eq.~(\ref{eq:probKGeneral}) reduces to 
\begin{eqnarray} \label{eq:probGTR}
\lefteqn{\Pr\left(\mathbf{L} = \boldsymbol\ell \mid \pi_i, C_{ij} \right)}  \qquad\qquad  \nonumber \\
& = & g(l_1, \ldots, l_K, l_{12}, \ldots, l_{K - 1, K} ; \pi_i, C_{ij}) \times \prod_{1 \le i < j \le K}h(\vec{l}_{ij}), 
\end{eqnarray}
where
\begin{eqnarray}
\lefteqn{g(l_1, \ldots, l_K, l_{12}, \ldots, l_{K - 1, K} ; \pi_i, C_{ij}) = \frac{L!}{(\prod_{i = 1}^K l_i!) (\prod_{1 \le i < j \le K} l_{ij}!)}}   \nonumber \\
 & & 				\quad \times	 \left(\prod_{i = 1}^K \left\{\pi_i - H_{M - 1} \sum_{j \ne i} C_{ij}\right\}^{l_i}  \right) 	\prod_{1 \le i < j \le K} (2 H_{M - 1}C_{ij})^{l_{ij}}, 
\end{eqnarray}
\begin{equation}
h(\vec{l}_{ij}) = \frac{1}{\textcolor{black}{{(2 H_{M - 1})}^{l_{ij}}}} \frac{l_{ij}!}{\prod_{y = 1}^{M - 1} l_{ij}(y)!}
	  \prod_{y = 1}^{M - 1} \left( \frac{1}{M - y} + \frac{1}{y} \right)^{l_{ij}(y)},  
\end{equation}

This  factorisation is a generalisation of that for the $K = 2$ rate matrix, which is necessarily reversible. The factors of $h(.)$ are again independent of the model parameters, and so $L_i$ and $L_{ij}$ form a set of sufficient statistics for estimating $\pi_i$ and $C_{ij}$.  
Furthermore, since by analogy with Eq.~(\ref{sumOverYGives1})
\begin{equation} 
\sum_{\{\vec{l}_{ij}: \,\sum_{y = 1}^{M - 1} l_{ij}(y) = l_{ij}\}} h(\vec{l}_{ij}) = 1, 
\end{equation}
the marginal probability in $L_i$ and $L_{ij}$ is a multinomial 
\begin{equation}	\label{marginalInL_iL_ij}
\Pr\left(\textcolor{black}{L_1 = l_1, ..., L_K =l_K}, L_{ij} = l_{ij} \mid \pi_i, C_{ij}  \right) = g(l_1, \ldots, l_K, l_{12}, \ldots, l_{K - 1, K} ; \pi_i, C_{ij}),  
\end{equation}  
with the same number of categories as the number of parameters to be estimated plus one. Again we have a flat family of multinomials and Eq.~(\ref{unbiasedPHat}) implies that $\hat{p}_{ij} = L_i/L$ and $\hat{p}_{ij} = L_{ij}/L$ are minimum variance, unbiased, ML estimators of  $p_i := \pi_i - H_{M - 1} \sum_{j \ne i} C_{ij}$ and $p_{ij} := 2 H_{M - 1}C_{ij}$ respectively. The required ML estimators assuming a {\em reversible} model rate matrix are then 
\begin{equation}	\label{estimatorsGTR}
\hat{C}_{ij} = \frac{L_{ij}}{2LH_{M - 1}}, \quad  \hat{\pi}_i= \frac{L_i + \frac{1}{2}\sum_{j \ne i}L_{ij}}{L}.
\end{equation}
By linearity these estimators are unbiased.

In \citet[][p~28]{BurdenTang17} it is incorrectly stated that Eq.~(\ref{estimatorsGTR}) are unbiased, ML estimators for the general {\em non-reversible} model. The mistake arose because of an incorrect use of the Neyman factorisation theorem: By an analogous argument to that used in Section~\ref{sec:MaxLikelihoodEstimatesKGeneral} above, the $C_{ij}$ cannot be decoupled from the $\Phi_{ij}$, and dividing the likelihood function Eq.(\ref{eq:probKGeneral}) by the marginal distribution, Eq.~(\ref{marginalInL_iL_ij}), does not give a quotient which is independent of the $C_{ij}$.  While, by Eq.~(\ref{unbiasedPHat}), it may be the case that Eq.~(\ref{estimatorsGTR}) are unbiased they are not the ML estimators for a general non-reversible model.  
%
%%%%%%%%%%%%%%%%%%%%%%%%%%%%%%%%%%%%%%%%%%%%%%%%%%%%%%%%%%%%
%
\section{Strand symmetry}
\label{sec:strandSymmetry}

\begin{figure}[t!]
\begin{center}
\centerline{\includegraphics[width=0.8\textwidth]{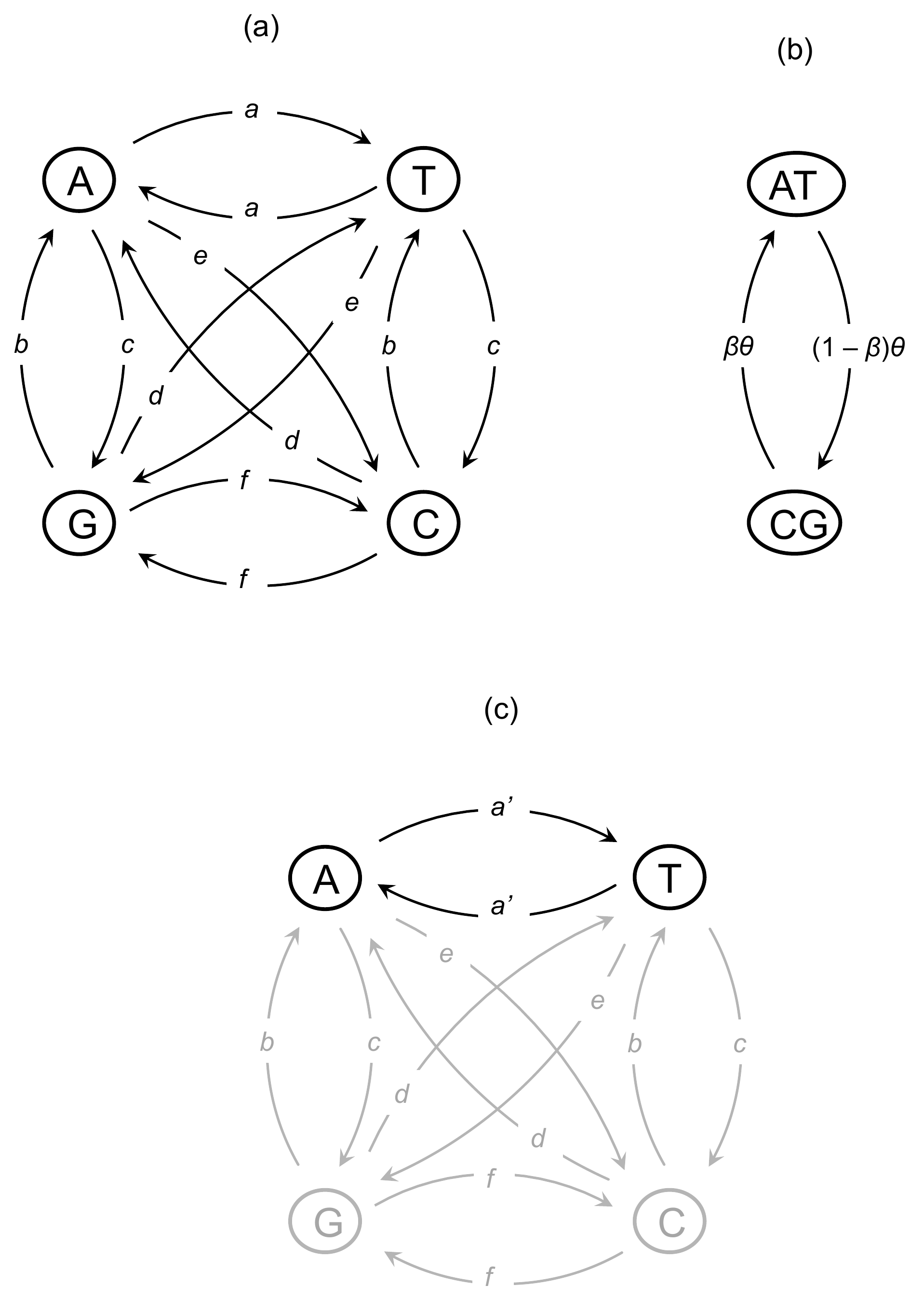}}
\caption{(a) The rate matrix with strand symmetry, Eq.~(\ref{strand-symmetric Q}), (b) the effective 2-allele rate matrix with alleles grouped as $(AG)$ and $(CG)$, Eq.~(\ref{eq:effective_2_allelle_Q}), and (c) the effective rate matrix conditioning on the event that a site is occupied by only $A$ and/or $T$ alleles, Eq.~(\ref{aPrimeQmatrix}). } 
\label{fig:Qmatrices}
\end{center}
\end{figure}

Most genomic sequences, when examined on a sufficiently large scale, are observed to be strand-symmetric, that is symmetric under simultaneous 
interchange of nucleotides $A$ with $T$ and $C$ with $G$~\citep{Baisnee02}.  
Any strand-symmetric rate matrix can be parameterised as shown in Fig.~\ref{fig:Qmatrices}(a) ~\citep{Lobry95}: 
\begin{equation}\label{strand-symmetric Q}
\mathbf{Q} = \begin{pmatrix}
 -a-c-e&  a&  c& e\\ 
 a&  -a-c-e& e & c\\ 
 b&  d& -b-d-f & f\\ 
 d& b &  f& -d-b-f
\end{pmatrix}  ,   
\end{equation}
where rows and columns are ordered $(A, T, G, C)$.  For the purposes of calculating ML estimates of $\mathbf{Q}$ from site frequency data, this parametrization turns out to be more convenient than the $(\pi_i, C_{ij}, \Phi_{ij})$ parameters introduced in Section~\ref{sec:QReparametrization} for more general rate matrices. The off-diagonal elements $a, \ldots, f$ are all assumed to be small.   

%
%%%%%%%%%%%%%%%%%%%%%%%%%%%%%%%%%%%%%%%%%%%%%%%%%%%%%%%%%%%%
%
\subsection{Stationary strand-symmetric sampling distribution}

The stationary distribution of the rate matrix Eq.~(\ref{strand-symmetric Q}) is 
\begin{equation}
\pi^{\rm T} = \tfrac{1}{2}(\begin{array}{cccc} \beta & \beta &  1 - \beta & 1 - \beta \end{array} ), 
\end{equation}
where 
\begin{equation}
\beta = \frac{b + d}{b + d + c + e}.  
\end{equation}
Combining Eqs.~(\ref{strand-symmetric Q}) and (\ref{eq:general_stationary_distribution}) gives the stationary sampling distribution to first order in the elements of $\mathbf Q$ as 
\begin{eqnarray}\label{eq:stationary_distribution_strand_symmetric}
\lefteqn{\Pr(\mathbf{Y} = \mathbf{y} \mid \mathbf{Q}) =} \nonumber \\
&& \begin{cases}
 \displaystyle \tfrac{1}{2}\{\beta(1-(a+c+e) H_{M-1})\} ,   & y_A=M,\quad y_{T,G,C}=0 \\
                                                                & \text{or } y_T=M,\quad y_{A,G,C}=0; \\ \\
 \displaystyle \tfrac{1}{2}\{(1-\beta)(1-(b+d+f)H_{M-1})\} ,  & y_G=M,\quad y_{A,T,C}=0; \\ 
                                                                & \text{or } y_C=M,\quad y_{A,T,G}=0; \\ 
 \displaystyle \tfrac{1}{2} \beta a \left\{\frac{1}{y_T} +\frac{1}{y_A}\right\} , & y_T+y_A=M, \, y_C = y_G = 0; \\ \\
 \displaystyle \tfrac{1}{2} \left\{\beta \frac{e}{y_C}+(1-\beta) \frac{d}{y_A}\right\} ,  &  y_C+y_A=M, \, y_T = y_G = 0; \\ \\
 \displaystyle \tfrac{1}{2} \left\{\beta \frac{e}{y_G}+(1-\beta) \frac{d}{y_T}\right\} ,  &  y_G+y_T=M, \, y_C = y_A = 0: \\ \\
 \displaystyle \tfrac{1}{2} \left\{\beta \frac{c}{y_G}+(1-\beta) \frac{b}{y_A}\right\} ,  &  y_G+y_A=M, \, y_C = y_T = 0; \\ \\
 \displaystyle \tfrac{1}{2} \left\{\beta \frac{c}{y_C}+(1-\beta) \frac{b}{y_T}\right\} ,  &  y_C+y_T=M, \, y_G = y_A = 0; \\ \\
 \displaystyle \tfrac{1}{2}(1-\beta) f \left\{\frac{1}{y_G} +\frac{1}{y_C}\right\} ,  &    y_G+y_C=M, \, y_T = y_A = 0; \\ \\
 0, & y_k, y_l, y_m >0 \text{ for distinct}\\ &\text{$k$, $l$, $m$.}
\end{cases}
\end{eqnarray}

In Section~\ref{sec:MaxLikelihoodEstimates} we isolated sufficient statistics within site frequency data for estimating certain combinations of parameters of $\mathbf{Q}$ by factoring the likelihood Eq.~(\ref{eq:likelihood}). Here we take a slightly different approach and carry out the factorisation at the level of the sampling distribution. This approach is equivalent in that it ultimately relies on properties of the multinomial distribution behind the likelihood, but is more suited to the strand-symmetric rate matrix. Define the reparametrization
\begin{equation} \label{eq:primedParameters}
\begin{split}
\beta &= \frac{b + d}{\theta}, \\
\gamma &= \frac{(b + d)(c + e)}{b + d + c + e} = \beta(1 - \beta)\theta, \\
a' &= \frac{\beta a}{\beta - \gamma H_{M - 1}}, \\
b' &= \frac{b}{b + d}, \\
e' &= \frac{e}{c + e}, \\
f' &= \frac{(1 - \beta) f}{1 - \beta - \gamma H_{M - 1}}, 
\end{split}
\end{equation}
where
\begin{equation}
\theta = b + c + d + e. 
\end{equation}
In terms of the six new parameters, the distribution Eq.~(\ref{eq:stationary_distribution_strand_symmetric}) factorises as 
\begin{eqnarray}\label{eq:stationary_distribution_strand_symmetric_reparam}
\lefteqn{\Pr(\mathbf{Y} = \mathbf{y} \mid \mathbf{Q}) =} \nonumber \\
&&\begin{cases}
   \tfrac{1}{2}(\beta - H_{M-1}\gamma)(1-H_{M-1}a') , &(y_A=M,y_{T,G,C}=0)\text{ or}\\
   &(y_T=M,y_{A,G,C}=0); \\ \\
   \tfrac{1}{2}((1-\beta) - H_{M-1}\gamma)(1-H_{M-1}f') , &(y_C=M,y_{A,T,G}=0)\text{ or}\\
   &(y_G=M,y_{A,T,C}=0);\\
   \tfrac{1}{2}\displaystyle(\beta - H_{M-1}\gamma)a'\left(\frac{1}{y_T} + \frac{1}{y_A}\right) , & 1\leq y_T,y_A\leq M-1,\\ 
   &y_T+y_A=M;\\
    \tfrac{1}{2}\displaystyle\gamma\left(\frac{e'}{y}+\frac{1-b'}{M-y}\right) ,  & (y_C = y, y_A = M - y)\text{ or }\\
   																		   & (y_G = y, y_T = M - y);\\ \\
    \tfrac{1}{2}\displaystyle\gamma\left(\frac{1-e'}{y}+\frac{b'}{M-y}\right) ,  & (y_G = y, y_A = M - y)\text{ or }\\
   																		   & (y_C = y, y_T = M - y);\\ \\
   \tfrac{1}{2}\displaystyle(1-\beta - H_{M-1}\gamma) f'\left(\frac{1}{y_G} + \frac{1}{y_C}\right) , & 1\leq y_G,y_C\leq M-1,\\ 
   &y_G+y_C=M\,  ; \\ \\
 0, & y_k, y_l, y_m >0 \text{ for distinct}\\ &\text{$k$, $l$, $m$,}
\end{cases}
\end{eqnarray}
correct to first order in the elements of $\mathbf Q$.

The motivation for the choice of $\beta$ and $\gamma$ in Eq.(\ref{eq:primedParameters}) comes from partitioning the genomic alphabet into two effective alleles, $(AT)$ and $(CG)$.  Following the procedure described in Appendix~B of \citet{BurdenTang16} the effective 2-allele model corresponds to the $2 \times 2$ rate matrix (see Fig.~\ref{fig:Qmatrices}(b)) 
\begin{equation} \label{eq:effective_2_allelle_Q}
\widetilde{\mathbf{Q}} = \left(\begin{array}{cc}
Q_{(AT),(AT)} & Q_{(AT),(CG)} \\
Q_{(CG),(AT)} & Q_{(CG),(CG)} 
\end{array} \right) 
= \theta \left(\begin{array}{cc}
\beta - 1 & 1 - \beta \\
    \beta  &    -\beta  
\end{array} \right), 
\end{equation}
whose stationary distribution is 
\begin{equation} 
\widetilde{\pi}^{\rm T} = \left( \begin{array}{cc} \beta & 1 - \beta \end{array} \right).  
\end{equation}
Note that the equivalence to a 2-allele model only extends to the stationary distribution as, in general, there is no 2-state Markov chain dynamically equivalent to the partitioning of a given multi-state Markov chain.  The stationary sampling distribution of a diffusion-limit 2-allele mutation-drift model is solved in Section~4 of \citet{Vogl14}.  Translated to the notation of the current paper, Vogl's result is 
\begin{eqnarray} \label{eq:sampling_distribution_2_allele}
\lefteqn{\Pr(Y_A + Y_T = y, Y_C + Y_G = M - y \mid \mathbf{Q} )} \nonumber \\ 
& = &  \Pr(\widetilde{Y}_{(AT)} = y, \widetilde{Y}_{(CG)} = M - y \mid \widetilde{\mathbf{Q}} ) \nonumber \\ 
& = &  \begin{cases}
\beta - \gamma H_{M - 1} +\mathcal{O}(\theta^{2}),                                                     &  y = M; \\
\displaystyle\gamma \left(\frac{1}{y} + \frac{1}{M - y}\right) +\mathcal{O}(\theta^{2}) , & 1 \le y \le M - 1;\\
1 - \beta - \gamma H_{M - 1} +\mathcal{O}(\theta^{2}),                                                & y = 0.
\end{cases}
\end{eqnarray}
It is straightforward to check that Eq.~(\ref{eq:sampling_distribution_2_allele}) is indeed the marginal distribution of Eq.~(\ref{eq:stationary_distribution_strand_symmetric_reparam}).  

The motivation for the choice of $a'$ in Eq.~(\ref{eq:primedParameters}) comes from conditioning on the event that the sampled site is occupied by only $A$ or $T$ alleles (see Fig.~\ref{fig:Qmatrices}(c)).  From Eq,~(\ref{eq:stationary_distribution_strand_symmetric}) and the definition of $\gamma$, we have
\begin{equation}	\label{eq:probSiteIsAT}
\Pr(Y_A + Y_T = M \mid \mathbf{Q} ) = \beta(1 - (c + e)H_{M - 1}) = \beta - \gamma H_{M - 1}.  
\end{equation}
The conditional probability that the site is occupied by $y$ $A$-alleles and $(M - y)$ $T$-alleles is then 
\begin{eqnarray}
\lefteqn{\Pr(Y_A = y \mid Y_A + Y_T = M, \mathbf{Q})  } \quad \nonumber \\
& = & 		\frac{\Pr(Y_A = y, Y_T = M - y \mid \mathbf{Q})}{\Pr(Y_A + Y_T = M, \mid \mathbf{Q})}\nonumber \\
& = & \begin{cases}
	\displaystyle \frac{\tfrac{1}{2}\beta(1 - (a + c + e)H_{M - 1})}{\beta - \gamma H_{M - 1}}, & y = 0 \text{ or } y = M; \\
	\displaystyle \frac{\tfrac{1}{2}\beta a}{\beta - \gamma H_{M - 1}}\left(\frac{1}{y} + \frac{1}{M - y}\right), & y = 1, \ldots, M - 1 
	\end{cases}	\nonumber \\
& = & \begin{cases}
	\displaystyle \tfrac{1}{2}\left(1 - \frac{\beta a}{\beta - \gamma H_{M - 1}}H_{M - 1}\right), & y = 0 \text{ or } y = M; \\
	\displaystyle \tfrac{1}{2}\frac{\beta a}{\beta - \gamma H_{M - 1}}\left(\frac{1}{y} + \frac{1}{M - y}\right), & y = 1, \ldots, M - 1 .
	\end{cases}
\end{eqnarray}
With the definition of $a'$ in Eq.~(\ref{eq:primedParameters}) this simplifies to 
\begin{equation}\label{eq:sampling_distribution_aPrime}
\Pr(Y_A = y \mid Y_A + Y_T = M, \mathbf{Q}) = \begin{cases} 
	\tfrac{1}{2} (1 - a'H_{M - 1}) , & y = 0 \text{ or } y = M; \\
	\tfrac{1}{2} a' \displaystyle\left(\frac{1}{y} + \frac{1}{M - y}\right), & y = 1, \ldots, M - 1.
	\end{cases}
\end{equation} 
Comparing with Eq.~(\ref{eq:sampling_distribution_2_allele}), it is clear this is the sampling distribution for an effective 2-allele mutation-drift model with rate matrix 
\begin{equation} \label{aPrimeQmatrix}
\tilde{\mathbf{Q}}^{(AT)} = a' \left(\begin{array}{rr}
	-1 &  1 \\
	 1 & -1 
	\end{array} \right).
\end{equation}
This interpretation is evident in the first and third lines of Eq.~(\ref{eq:stationary_distribution_strand_symmetric_reparam}).  
An analogous argument holds for the choice of parameter $f'$ by conditioning on the event that the sampled site is occupied by only $C$ or $G$ alleles.    
%
%%%%%%%%%%%%%%%%%%%%%%%%%%%%%%%%%%%%%%%%%%%%%%%%%%%%%%%%%%%%
%
\subsection{Strand-symmetric parameter estimation}
\label{sec:SSEstimation}

Assume a dataset in the form of a site frequency spectrum obtained by sampling $L$ independent neutrally evolving sites within a multiple alignment of $M$ genomes with allele occupancy counts defined as in Eqs.~(\ref{eq:L_definitions1}) and (\ref{eq:L_definitions2}). Also define 
\begin{equation}    \label{eq:L_definitionsSS}
\begin{split}
L^{(AT)} &= L_A + L_T + L_{AT} \\
L^{(CG)} &= L_C + L_G + L_{CG} \\
L^{(AT,CG)} &= L_{AC} + L_{AG} + L_{TC} + L_{TG}.  
\end{split}
\end{equation}
Since Eq.~(\ref{eq:general_stationary_distribution}) implies that tri-allelic and tetra-allelic sites only occur with probability ${\mathcal O}(\theta^2)$, as in Section~\ref{sec:siteFreqData} assume that 
\begin{equation}
L = \sum_i L_i +L_{\rm P} = L^{(\rm AT)} + L^{(\rm CG)} + L^{(\rm AT,CG)}.    
\end{equation}

From Eq.~(\ref{eq:stationary_distribution_strand_symmetric_reparam}) we obtain the likelihood function 
\begin{eqnarray} \label{eq:likelihoodSS}
\lefteqn{{\cal L}(\beta, \gamma,  a', b', e' f' \mid l_i, l_{ij}(y)) = \Pr\left(L_i = l_i, L_{ij}(y) = l_{ij}(y) \mid \mathbf{Q} \right)} \nonumber \\
& = & \frac{L!}{(\prod_i l_i!) (\prod_{i < j} \prod_{y = 1}^{M - 1} l_{ij}(y)!)} \times \left\{ \tfrac{1}{2}(\beta - H_{M-1}\gamma)(1-H_{M-1}a') \right\}^{l_A + l_T} \nonumber \\
&  & \qquad \times \left\{ \tfrac{1}{2}(1 - \beta - H_{M-1}\gamma)(1-H_{M-1}f') \right\}^{l_C + l_G} \nonumber \\
& & \qquad \times \left\{ \tfrac{1}{2}(\beta - H_{M-1}\gamma)a' \right\}^{L_{AT}} \prod_{y=1}^{M - 1} \left( \frac{1}{M - y} + \frac{1}{y} \right)^{l_{AT}(y)} \nonumber\\
& & \qquad \times \left\{ \tfrac{1}{2}(1 - \beta - H_{M-1}\gamma)f' \right\}^{L_{CG}} \prod_{y=1}^{M - 1} \left( \frac{1}{M - y} + \frac{1}{y} \right)^{l_{CG}(y)} \nonumber\\
& & \qquad \times \left(\tfrac{1}{2}\gamma\right)^{L_{AC} + L_{TG}} \prod_{y=1}^{M - 1} \left( \frac{e'}{M - y} + \frac{1 - b'}{y} \right)^{l_{AC}(y) + l_{TG}(y)} \nonumber\\
& & \qquad \times \left(\tfrac{1}{2}\gamma\right)^{L_{AG} + L_{TC}} \prod_{y=1}^{M - 1} \left( \frac{1 - e'}{M - y} + \frac{b'}{y} \right)^{l_{AG}(y) + l_{TC}(y)} \nonumber\\
& = & \text{(const.)}\times (2 H_{M - 1}\gamma)^{l^{(AT,CG)}}  \nonumber \\
& & \qquad \times \left\{(1 - H_{M-1}\gamma) H_{M - 1} a' \right\}^{l_{AT}}  \nonumber \\
& & \qquad \times \left\{(1 - H_{M-1}\gamma) (1 - H_{M - 1} a') \right\}^{l_A + L_T}  \nonumber \\
& & \qquad \times \left\{(1 - \beta - H_{M-1}\gamma) H_{M - 1} f' \right\}^{l_{CG}}  \nonumber \\
& & \qquad \times \left\{(1 - \beta - H_{M-1}\gamma) (1 - H_{M - 1} f') \right\}^{l_C + l_G}  \nonumber \\
& & \qquad \times \prod_{y=1}^{M - 1} \left\{ \left( \frac{e'}{M - y} + \frac{1 - b'}{y} \right)^{l_{AC}(y) + l_{TG}(y)} 
								   \left( \frac{1 - e'}{M - y} + \frac{b'}{y} \right)^{l_{AG}(y) + l_{TC}(y)} \right\},  \nonumber\\
\end{eqnarray}
where the constant is a combinatorial factor independent of $\beta$, $\gamma$,  $a'$, $b'$, $e'$, and $f'$. This is a multinomial distribution, which can be factored in two different ways.  

Firstly, \textcolor{black}{recall the notation $\vec{l}_{ij}$ defined by Eq.~(\ref{vecLDefn})} and 
consider 
\begin{eqnarray} \label{firstFactorisationSS}
\lefteqn{{\cal L}(\beta, \gamma,  a', b', e' f' \mid l_i, l_{ij}(y)) = g(l^{(AT)}, l^{(CG)}, l^{(AT, CG)}; \beta, \gamma)} \nonumber \\
&  & \qquad \times h(l_A, l_T, l_C, l_G, l_{AT}, l_{CG}, \vec{l}_{AC}, \vec{l}_{TG}, \vec{l}_{AG}, \vec{l}_{TC}; a', f', b' e'), 
\end{eqnarray}
where 
\begin{eqnarray} 
\lefteqn{g(l^{(AT)}, l^{(CG)}, l^{(AT, CG)}; \beta, \gamma) =  \frac{L!}{l^{(AT)})! l^{(CG)}! l^{(AT,CG)}!}} \nonumber \\
&  & \quad \times (\beta - H_{M-1}\gamma)^{l^{(AT)}} (1 - \beta - H_{M-1}\gamma)^{l^{(CG)}} (2 H_{M - 1}\gamma)^{l^{(AT,CG)}}, 
\end{eqnarray}
and the function $h(\cdot)$ is a product of the remaining factors times an appropriate combinatorial factor to ensure that $\cal L$ is correctly normalised. Following the same procedure as for the general rate matrix, observe that $h(\cdot)$ is independent of $\beta$ and $\gamma$, and that Neyman factorisation then implies that $L^{(AT)}$, $L^{(CG)}$ and $L^{(AT,CG)}$ are sufficient statistics for estimating $\beta$ and $\gamma$.  Summing over the redundant allele occupancy counts subject to conditioning on $L^{(AT)} = l^{(AT)}$, $L^{(CG)} = l^{(CG)}$ and $L^{(AT,CG)} = l^{(AT,CG)}$ gives the flat family of trinomials $g(l^{(AT)}, l^{(CG)}, l^{(AT, CG)}; \beta, \gamma)$, from which we read off the minimum variance, unbiased, ML estimators 
\begin{equation}
\widehat{\beta - H_{M-1}\gamma} = \frac{L^{(AT)}}{L}, \qquad \widehat{2H_{M-1}\gamma} = \frac{L^{(AT,CG)}}{L}.  
\end{equation}
By linearity we therefore have that 
\begin{equation}		\label{betaGammaEstimators}
\hat{\beta} = \frac{L^{(AT)} + \tfrac{1}{2}L^{(AT,CG)}}{L}, \qquad \hat{\gamma} = \frac{L^{(AT,CG)}}{2LH_{M - 1}}, 
\end{equation}
are unbiased ML estimators. These estimators agree precisely with corresponding estimators derived by \citet[][Section~4.1]{Vogl14} and re-derived in Section~\ref{sec:MaxLikelihoodEstimatesKEquals2} for the 2-allele mutation-drift model with stationary sampling distribution equivalent to Eq.~(\ref{eq:sampling_distribution_2_allele}).  

Secondly, consider the factorisation 
\begin{eqnarray} \label{secondFactorisationSS}
\lefteqn{{\cal L}(\beta, \gamma,  a', b', e' f' \mid l_i, l_{ij}(y))}  \nonumber \\
& = & g(l_A + l_T, l_C + l_G, l_{AT}, l_{CG}, l^{(AT, CG)}; \beta, \gamma, a', f') \nonumber \\
&  & \qquad \times h(\vec{l}_{AC}, \vec{l}_{TG}, \vec{l}_{AG}, \vec{l}_{TC}; b' e'), 
\end{eqnarray}
where 
\begin{eqnarray}	\label{gSecondFactorisationSS}
\lefteqn{g(l_A + l_T, l_C + l_G, l_{AT}, l_{CG}, l^{(AT, CG)}; \beta, \gamma, a', f')} \nonumber \\
& = & \frac{L!}{l^{(AT, CG)}! l_{AT}! (l_A + l_T)! l_{CG}! (l_C + l_G)!}  \nonumber \\
& & \times (2 H_{M - 1}\gamma)^{l^{(AT,CG)}} {p_{AT}}^{l_{AT}} {p_{A + T}}^{l_A + l_T} {p_{CG}}^{l_{CG}} {p_{C + G}}^{l_C + l_G}, 
\end{eqnarray}
with 
\begin{equation}
\begin{split}
p_{AT} &= (\beta - H_{M - 1}\gamma)H_{M-1}a' = H_{M - 1}\beta a\\
p_{A + T} &= (\beta - H_{M - 1}\gamma)(1-H_{M-1}a')\\
p_{CG} &= (1 - \beta - H_{M - 1}\gamma)H_{M-1}f' = H_{M - 1}(1 -\beta) f\\
p_{C + G} &= (1 - \beta - H_{M - 1}\gamma)(1-H_{M-1}f')\\
\end{split}
\end{equation}
and $h(\vec{l}_{AC}, \vec{l}_{TG}, \vec{l}_{AG}, \vec{l}_{TC}; b' e')$ is the final product in Eq.~(\ref{eq:likelihoodSS}) times an appropriate combinatorial factor to ensure that $\cal L$ is correctly normalised. Applying Neyman factorisation as before to factor out $h(.)$, and recognising Eq.~(\ref{gSecondFactorisationSS}) as a flat family of fifth order multinomials in 4 independent parameters, we obtain the minimum variance, unbiased, ML estimators  
$\hat{p}_{AT} = L_{AT}/L$ and $\hat{p}_{CG} = L_{CG}/L$.  Thus 
\begin{equation}	\label{AFEstimators}
\begin{split} 
\widehat{\beta a} = \frac{L_{AT}}{L H_{M - 1}}, \\
\widehat{(1 - \beta)f}  = \frac{L_{CG}}{L H_{M - 1}}, 
\end{split}
\end{equation}
are minimum variance, unbiased ML estimators.  

There is no simple analytic formula for the ML estimators $\widehat{b'}$ and $\widehat{e'}$.  However they can be easily computed by maximising the conditional log-likelihood arising from $h(\vec{l}_{AC}, \vec{l}_{TG}, \vec{l}_{AG}, \vec{l}_{TC}; b' e')$, namely 
\begin{eqnarray}\label{eq:cond_ll_eb}
\log {\cal L}(e', b') & = & \sum_{y = 1}^{M - 1} \left\{(l_{AC}(y) + l_{TG}(y)) \log\left( \frac{e'}{M - y} + \frac{1 - b'}{y} \right) \right. \nonumber \\ 
			&  &    \quad   \left. +\, (l_{AG}(y) + l_{TC}(y)) \log\left( \frac{1 - e'}{M - y} + \frac{b'}{y} \right) \right\} 
\end{eqnarray}
over the region $(b', e') \in [0, 1] \times [0, 1]$.  This can be done using, for instance, the R function {\tt  constrOptim( )} or the EM algorithm described in \ref{section:EM_algorithm}. The maximum is unique by the following argument:  
Set $\ell_1(y) = l_{AC}(y) + l_{TG}(y)$, $\ell_2(y) = l_{AG}(y) + l_{TC}(y)$ and 
$$
G_1(y) = \frac{e'}{M - y} + \frac{1 - b'}{y}, \qquad G_2(y) = \frac{1 - e'}{M - y} + \frac{b'}{y}.  
$$
Assuming $\min_y(\ell_1(y), \ell_2(y)) > 0$, the Hessian matrix
\begin{eqnarray}
{\cal H} & = & \left( \begin{array}{cc} 
\dfrac{\partial^2 \log {\cal L}}{\partial e'^2} & \dfrac{\partial^2 \log {\cal L}}{\partial e' \partial b'} \\ \\
\dfrac{\partial^2 \log {\cal L}}{\partial e' \partial b'} & \dfrac{\partial^2 \log {\cal L}}{\partial b'^2} 
\end{array} \right)		\nonumber \\
& = &\sum_{y = 1}^{M - 1} \left(\frac{\ell_1}{G_1^2} + \frac{\ell_2}{G_2^2}\right) 
\left( \begin{array}{cc} 
\dfrac{-1}{(M - y)^2} & \dfrac{1}{y(M - y)} \\ \\
\dfrac{1}{y(M - y)} & \dfrac{-1}{y^2} 
\end{array} \right), 
\end{eqnarray}
is negative definite since for any real $\mathbf{u}^{\rm T} = (u_1, u_2)$, 
\begin{equation}
\mathbf{u}^{\rm T} {\cal H} \mathbf{u} = - \sum_{y = 1}^{M - 1} \left(\frac{\ell_1(y)}{G_1(y)^2} + \frac{\ell_2(y)}{G_2(y)^2}\right) \left(\frac{u_1}{M - y} - \frac{u_2}{y}\right)^2 < 0.  
\end{equation}
Therefore any stationary point in the connected region for which $\log {\cal L}(e', b')$ is real must be an isolated local maximum.  More than one maximum cannot occur 
without there being a saddle point on a curve connecting them, so the maximum is unique.  

To obtain ML estimators of the strand-symmetric Q parameters $(a, \ldots, e)$ in Eq.~(\ref{strand-symmetric Q}), we invert 
Eq.~(\ref{eq:primedParameters}) to give 
\begin{equation} \label{eq:invertPrimedParameters}
\begin{split}
b &= b' \frac{\gamma}{1 - \beta}, \\
c &= (1 - e')\frac{\gamma}{\beta}, \\
d &= (1 - b')\frac{\gamma}{1 - \beta}, \\
e &= e' \frac{ \gamma}{\beta}, \\
\end{split}
\end{equation}
Then from Eqs.~(\ref{betaGammaEstimators}), (\ref{AFEstimators}) and (\ref{eq:invertPrimedParameters}) we obtain the ML estimators  
\begin{equation} \label{eq:unprimedEstimators}
\begin{split}
\hat a &= \frac{1}{H_{M - 1}} \frac{L_{AT}}{L^{(AT)} + \tfrac{1}{2} L^{(AT, CG)}}, \\
\hat b &= \frac{\widehat{b'}}{2H_{M - 1}} \frac{L^{(AT, CG)}}{L^{(CG)} + \tfrac{1}{2} L^{(AT, CG)}}, \\
\hat c &= \frac{1 - \widehat{e'}}{2H_{M - 1}} \frac{L^{(AT, CG)}}{L^{(AT)} + \tfrac{1}{2} L^{(AT, CG)}}, \\
\hat d &= \frac{1 - \widehat{b'}}{2H_{M - 1}} \frac{L^{(AT, CG)}}{L^{(CG)} + \tfrac{1}{2} L^{(AT, CG)}}, \\
\hat e &= \frac{\widehat{e'}}{2H_{M - 1}} \frac{L^{(AT, CG)}}{L^{(AT)} + \tfrac{1}{2} L^{(AT, CG)}}, \\
\hat f &= \frac{1}{H_{M - 1}} \frac{L_{CG}}{L^{(CG)} + \tfrac{1}{2} L^{(AT, CG)}}. 
\end{split}
\end{equation}
Note that these estimators are in general biased.  
%
%%%%%%%%%%%%%%%%%%%%%%%%%%%%%%%%%%%%%%%%%%%%%%%%%%%%%%%%%%%%
%
\section{Application}
\label{sec:Application}

\citet{BergmanBetancourtVogl18} extracted sequence information of 197 {\em Drosophila melanogaster} individuals \citep{Lack15} on the short autosomal introns (i.e.\ the nucleotides in positions $8$ through $30$ of introns $\leq 65$ bp in length), resulting in a site frequency spectrum of 218,942 nucleotides. This dataset is one of the largest and most accurate available today. As the population of {\em D.\ melanogaster} does not seem to be in equilibrium 
but instead exhibits a bias towards $C$ and $G$ nucleotides, the data are nevertheless not ideal for our purpose. 

We implemented our ML estimators in the statistical programming 
language R \citep{RCoreTeam17}.  
The ML estimate of the general rate matrix $\mathbf Q$ was determined 
using Eqs.~(\ref{C_piPrime_Estimates}) to estimate $\pi_1'$, $\pi_2'$,
$\pi_3'$ and $C$, and the R function {\tt constrOptim()} to maximise 
$h(\boldsymbol\ell_{\rm P}; c_{ij}, \phi_{ij})$ defined by
Eq.~(\ref{hGeneralReparam}) with respect to the set of independent 
parameters $c_{ij}$ and $\phi_{ij}$ defined in Eq.~(\ref{independentParams}).  For $K = 4$ this amounts to five $c_{ij}$'s and three $\phi_{ij}$'s, 
i.e.\ a total of eight parameters to be determined numerically.  
\textcolor{black}{
To optimise performance of the program a rough approximation to the 
maximum likelihood estimates of $c_{ij}$ and $\phi_{ij}$ is first 
determined from the  simpler but 
incorrect method in \citet[][p~28]{BurdenTang17}, and this approximation 
is then used as a starting values in the function {\tt constrOptim()} 
for the full 8 parameter 
optimizaton.  With this strategy the maximum likelihood estimate for the 
full 12-parameter model was calculated correct to 5 significant figures in less than 1 second 
on a MacBook Air laptop computer with a 1.8 GHz Intel Core i5 processor.  
}
The estimate of the rate matrix reconstructed via Eqs.~(\ref{QinTermsOfCPhi}) and (\ref{c_phiDefns}) is  
\begin{equation}	\label{QHatGeneral}
\hat{\mathbf{Q}} = \left( \begin{array}{rrrr}
 -0.018072 &  0.006464 &  0.007479 &  0.004129 \\
  0.006215 & -0.017885 &  0.003640 &  0.008030 \\
  0.016583 &  0.007047 & -0.028496 &  0.004866 \\
  0.006590 &  0.015678 &  0.004466 & -0.026734
  \end{array} \right), 
\end{equation}
where rows and columns are ordered $(A, T, G, C)$.  

Similarly the ML estimator assuming reversibility was calculated from Eqs.~(\ref{QinTermsOfCPhi}) and (\ref{estimatorsGTR}) with $\hat\Phi_{ij} \equiv 0$ as 
\begin{equation}
\hat{\mathbf{Q}}^\text{rev} = \left( \begin{array}{rrrr}
 -0.018077 &  0.006452 &  0.007697 &  0.003928 \\
  0.006227 & -0.017882 &  0.003483 &  0.008173 \\
  0.015987 &  0.007495 & -0.028477 &  0.004996 \\
  0.007098 &  0.015302 &  0.004346 & -0.026747
  \end{array} \right). 
\end{equation}
The ML estimator assuming strand symmetry was calculated from Eqs.~(\ref{strand-symmetric Q}) and (\ref{eq:unprimedEstimators}), with 
$\widehat{b'}$ and $\widehat{e'}$ obtained by numerically maximising Eq.~(\ref{eq:cond_ll_eb}), to obtain 
\begin{equation}
\hat{\mathbf{Q}}^\text{strand-sym} = \left( \begin{array}{rrrr}
 -0.017978 &  0.006337 &  0.007760 &  0.003880 \\
  0.006337 & -0.017978 &  0.003880 &  0.007760 \\
  0.016099 &  0.006804 & -0.027552 &  0.004649 \\
  0.006804 &  0.016099 &  0.004649 & -0.027552
  \end{array} \right). 
\end{equation}

Although the three estimated rate matrices do not differ greatly, the differences enable us to quantify 
the significance of the deviation of the general model 
from reversibility and strand symmetry. If reversibility (resp.\ strand symmetry) is taken as a null 
hypothesis and the general rate matrix taken as the alternate 
hypothesis, the log of the likelihood ratio statistic,  
\begin{equation}	\label{logLRatio}
- 2 \left[ \log{\cal L}(\hat{\mathbf{Q}}^\text{null} \mid \boldsymbol\ell) - \log{\cal L}(\hat{\mathbf{Q}}^\text{alt} \mid \boldsymbol\ell) \right] ,
\end{equation}
will asymptotically (as the number of sites $L \rightarrow \infty$) have a chi-squared distribution 
with $d = 3$ (resp. $d = 6$) degrees of freedom if 
the extra parameters required to specify the general rate matrix are not significant.  
Setting $\hat{\mathbf{Q}}^\text{null} = \hat{\mathbf{Q}}^\text{rev}$ or $\hat{\mathbf{Q}}^\text{strand-sym}$ 
in Eqs.~(\ref{eq:likelihood}) and (\ref{logLRatio}) gives p-values of $3.6 \times 10^{-5}$ and $< 10^{-91}$ 
respectively, indicating significant deviations from both reversibility and strand symmetry for this dataset. 
Note that \citet{BergmanBetancourtVogl18} report slight but significant deviations from Chargaff's second parity rule. 

In all three models, estimates for scaled mutation rates are higher for transitions ($A \leftrightarrow G$ 
and $C \leftrightarrow T$) than for transversions (($A \text{ or } G) \leftrightarrow (C \text{ or } T$)), as expected.  
By considering a sample of size $M = 2$ in the sampling distribution Eq.~(\ref{eq:general_stationary_distribution}), 
the first order approximation to the expected heterozygosity is 
\begin{equation}
1 - \sum_i \hat\pi_i\left(1 - \sum_{j\neq i} \hat{Q}_{ij}\right) = - \sum _{i \in \{A, T, G, C\}} \hat\pi_i \hat{Q}_{ii}. 
\end{equation}
For all three models this gives an identical expected heterozygosity of 0.0212.  
%
%%%%%%%%%%%%%%%%%%%%%%%%%%%%%%%%%%%%%%%%%%%%%%%%%%%%%%%%%%%%
%
\section{\textcolor{black}{Simulation}}
\label{sec:Simulation}

\textcolor{black}{
To assess the accuracy of the small-$\theta$ approximation we have simulated datasets
of site frequency spectra corresponding to sampling of $M$ individuals 
from a finite population of size $N$ at $L$ independent genomic sites.  For a given
rate matrix $Q$ we begin by numerically determining the stationary distribution 
of the finite population Wright-Fisher model with neutral mutation rates
$u_{ij} = (\exp(Q/(2N)))_{ij}$, consistent with Eq.~(\ref{continuumScaling}).  
For this step the 
full ${N + K - 1\choose K - 1} \times {N + K - 1 \choose K - 1}$ Markov transition matrix 
is used, where $K = 4$ for the genomic alphabet $\{A, T, G, C\}$.  For each site 
within each dataset, the allele occupancies are determined by drawing from a multinomial 
distribution corresponding to $M$ trials from $K$ categories with probabilities 
of success in each category determined by a random draw from the Wright-Fisher 
stationary distribution.  For each dataset this leads to site frequency data as 
defined by Eq.~(\ref{eq:L_definitions1}) from which an estimate $\hat{Q}$ of the full 
rate matrix is obtained from Eqs.~(\ref{hGeneralReparam}) 
and (\ref{C_piPrime_Estimates}) as described in Section~\ref{sec:MaxLikelihoodEstimatesKGeneral}. 
}

\textcolor{black}{
In order to illustrate the limit of applicability of the small-$\theta$ approximation we 
consider two examples for the `true' rate matrix $Q$: (i) the matrix given by 
Eq.~(\ref{QHatGeneral}) obtained from the {\em Drosophila melanogaster} of the previous 
section, and (ii) the same matrix multiplied by $0.1$.   The population 
size, which is constrained by the computational demands of determining the stationary 
state of a large Markov transition matrix, is chosen to be $N = 40$.  
Note that in general the diffusion $N \to \infty$ limit is approached very rapidly for the 
stationary distribution of a neutral Wright-Fisher model (see for example
\citet[][Figs.~5 and 6]{BurdenTang16}.)  We have aimed to satisfy the ideal limit 
$1 << M << N$ within the constraints of the simulation by choosing the sample 
size $M = 10$.  The number of genomic sites is $L = 10^5$.  
}

\begin{figure}[t!]
\begin{center}
\centerline{\includegraphics[width=\textwidth]{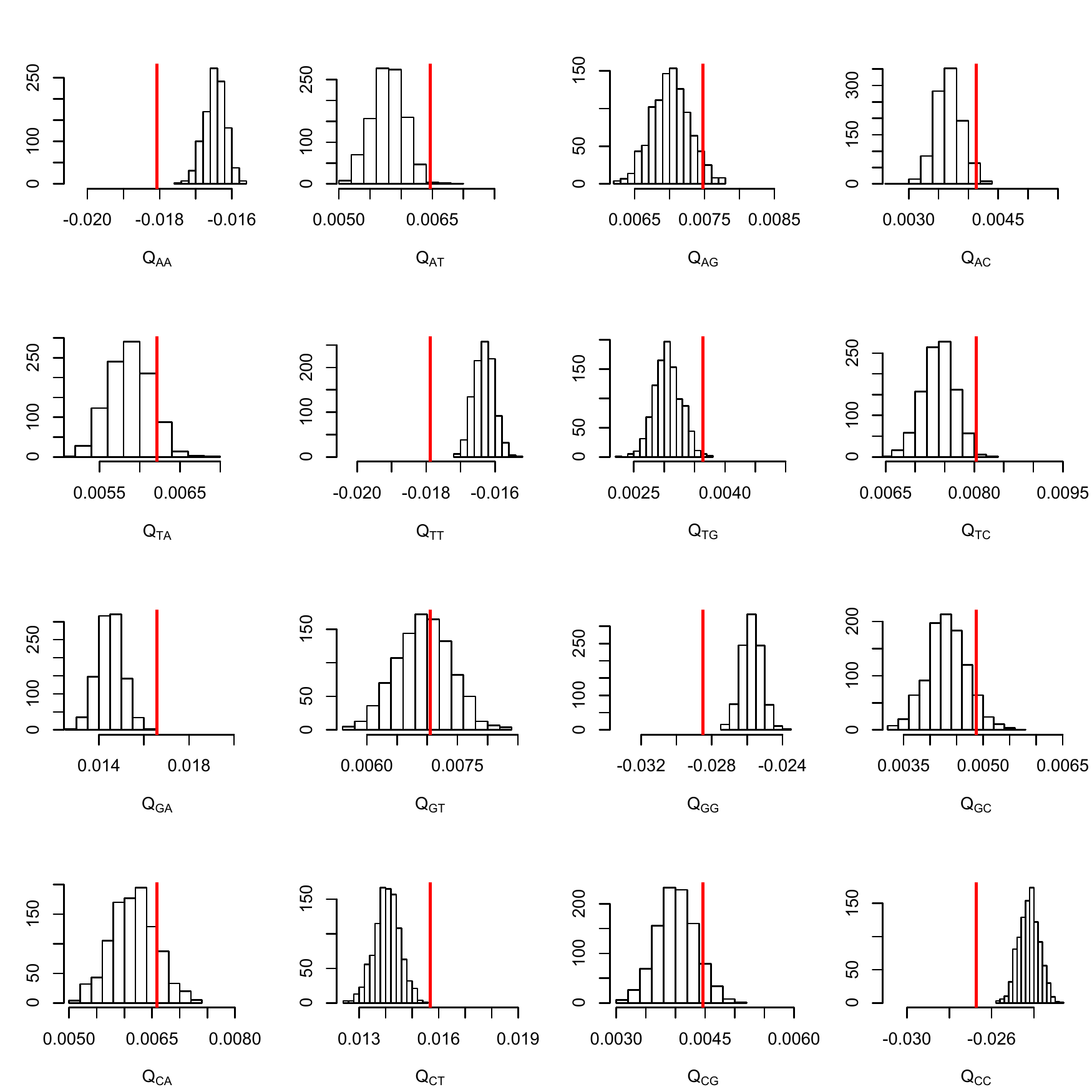}}
\caption{Histograms of estimates $\hat{Q}_{ij}$ of the rate matrix obtained from $1000$ 
synthetic datasets corresponding to sampling $M = 10$ individuals 
from a stationary-distribution population of $N = 40$ neutrally evolving haploid 
individuals at $L = 10^5$ genomic sites.  Elements of `true' rate matrix, 
Eq.~(\ref{QHatGeneral}), are shown as red vertical bars.} 
\label{fig:QHatHistograms}
\end{center}
\end{figure}

\begin{figure}[t!]
\begin{center}
\centerline{\includegraphics[width=\textwidth]{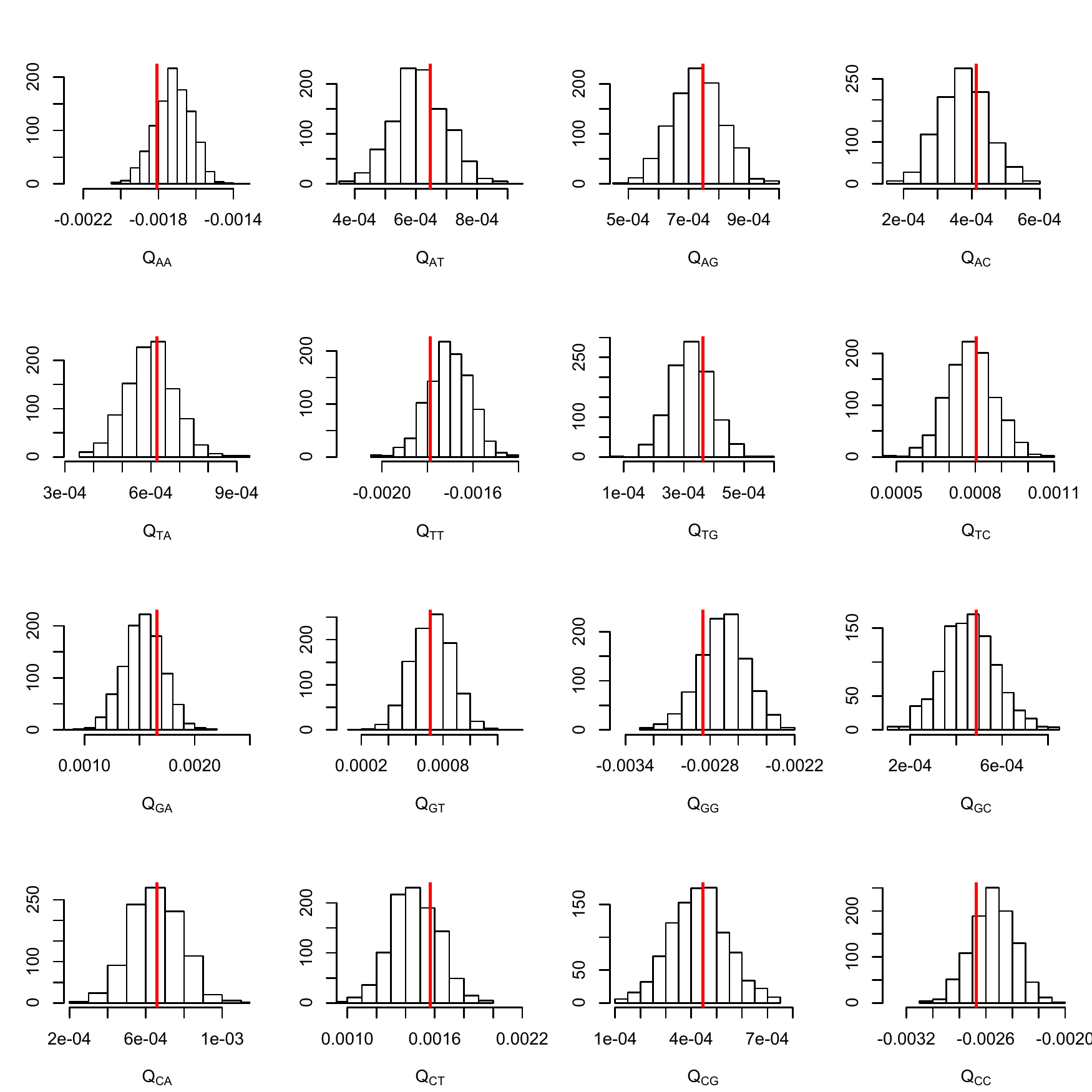}}
\caption{Same as Fig.~\ref{fig:QHatHistograms}, except that the `true' rate matrix 
is chosen to be $0.1$ times the matrix in Eq.~(\ref{QHatGeneral}).} 
\label{fig:QHatHistogramsSmallQ}
\end{center}
\end{figure}

\textcolor{black}{
Figures~\ref{fig:QHatHistograms} and \ref{fig:QHatHistogramsSmallQ} show histograms of the 
estimated rate matrix elements from $1000$ simulated datasets, 
together with the `true' matrix elements for the two rate matrices.  The off-diagonal 
elements of $Q$ are generally underestimated in Fig.~\ref{fig:QHatHistograms}
and, given that the diagonal elements are simply calculated as minus the sum of 
estimates of the off-diagonal row elements, the diagonal elements are 
correspondingly overestimated.  By comparison, the off-diagonal elements in 
Fig.~(\ref{fig:QHatHistogramsSmallQ}) are slightly underestimated, but generally 
acceptable.  The sum of the off-diagonal elements is $0.0912$ for the rate matrix 
in Fig.~\ref{fig:QHatHistograms} and $0.00912$ for the rate matrix in 
Fig.~\ref{fig:QHatHistogramsSmallQ}.  We conclude that, as a rule of thumb, 
the small-$\theta$ approximation will provide acceptable estimates provided  
the sum of the off-diagonal elements of the scaled rate matrix is less than 
$10^{-2}$.  Note that the widths of the histograms are determined by the sample size 
$M$, which is necessarily small in this simulation due to computational constraints, 
and are not indicative of standard errors expected in a biological dataset such 
as the {\em Drosophila melanogaster} dataset used in the previous section.  
}

% Figure fig:QHatHistograms

% Figure fig:QHatHistogramsSmallQ

%
%%%%%%%%%%%%%%%%%%%%%%%%%%%%%%%%%%%%%%%%%%%%%%%%%%%%%%%%%%%%
%
\section{Conclusions}
\label{sec:Conclusions}

We have obtained ML estimates of a scaled rate matrix from population allele frequencies observed in unlinked neutrally evolving, 
independent genomic sites under three sets of model assumptions: that the rate matrix is unconstrained, that it is reversible, 
and that it is strand-symmetric. The analysis is carried out to first order in scaled mutation rates $Q_{ij}$ 
defined by Eq.~(\ref{continuumScaling}), which are assumed to be small. This is equivalent to assuming that at most 
one mutation has occurred in the coalescent tree of the example, and hence that the sample includes only non-segregating and 
bi-allelic sites. The purpose of our analysis is twofold:  

Firstly, our treatment is more rigorous than a previous analysis of this problem by \citet[][p28]{BurdenTang17}, 
and corrects an error in that earlier analysis. The correct estimates, specifically for the case of a general 
unconstrained rate matrix and for the assumption of a strand-symmetric rate matrix, are given in Sections~\ref{sec:MaxLikelihoodEstimatesKGeneral} and \ref{sec:SSEstimation} respectively of the current paper. Although the correction is generally small in 
absolute terms, it is necessary for an accurate significance test of violation of reversibility and strand 
symmetry via the likelihood ratio statistic Eq.~(\ref{logLRatio}).  

Secondly, in Section~\ref{sec:Application} we have demonstrated efficient software in R implementing ML 
estimates for a biological dataset consisting of a site frequency spectrum extracted from short autosomal 
introns in a sample of {\em Drosophila melanogaster} individuals. This software is available at the web 
address given below, and requires as input a table of allele occupancy counts as defined by Eq.~(\ref{eq:L_definitions1}).  

It is worth stressing the limits on the use of our software for rate matrix estimation: The theory 
leading to the likelihood function Eq.~(\ref{eq:likelihood}) assumes a mutation-drift model corresponding 
to the backward generator Eq.~(\ref{backwardGenerator}), that is, the diffusion limit of a Moran or 
Wright-Fisher model for a population of constant size. Substitutions are assumed to be due to neutral 
mutations with no directional selection. Although \citet{VoglBergman15} have performed a similar ML 
analysis for the analogous bi-allelic model with selection parameters, we are unaware of any analytic 
solution for the multi-allelic sampling distribution with selection.  

The likelihood function is derived from the stationary sampling distribution. As pointed out in 
Section~\ref{sec:Application} the assumption of stationarity may not hold for our test 
{\em Drosophila melanogaster} dataset. The non-stationary sampling distribution has been derived to 
first order in $Q_{ij}$ by \citet{BurdenGriffiths19b}, and is considerably more complicated than 
the stationary distribution, Eq.~(\ref{eq:general_stationary_distribution}). Nevertheless it has the 
potential to serve as a basis for estimating neutral mutation rates in a non-stationary setting.  

The theory also assumes that the genomic loci should not only be neutrally evolving, but should 
have independent ancestries to avoid correlations due to common coalescent trees~\citep{SlatkinHudson91}. 
This should be possible in randomly mating diploid populations by choosing loci which are unlinked 
due to recombination. There is strong evidence that this requirement is satisfied for the 
{\em Drosophila melanogaster} dataset analysed in Section~\ref{sec:Application}~\citep{ClementeVogl12}.  

Finally, we stress that the first order analysis we have used assumes 
that all off-diagonal elements of $\mathbf{Q}$ are small. 
\textcolor{black}{
For situations where this is not the case it is necessary to resort to other approximation methods such as importance sampling 
\citep{StephensDonnelly00}, which, as noted by \citet[][p421--2]{DeIorio04}, are only expected to be exact for a parent-independent rate matrix,
in which case the sampling distribution is known to be multinomial-Dirichlet.  Application of importance sampling to the related problem 
of a mutations in a subdivided population with high mutation and migration rates have unfortunately proved to be considerably more computationally 
intensive than computations in the current paper \citep{DeIorio05}.  
}

\textcolor{black}{
Numerical simulations in Section~\ref{sec:Simulation} indicate that the small $\theta$ approximation will 
provide acceptable results provided the sum of 
the off-diagonal elements of $\mathbf{Q}$ are less than $10^{-2}$.  While the {\em Drosophila melanogaster} 
short intron dataset used in Section~\ref{sec:Application} 
appears to be just beyond the limit of of this restriction, the condition is believed to be satisfied in 
general for silent sites in protein coding genes in eukaryote 
genomes~\citep{Lynch16}.  
}
%
%%%%%%%%%%%%%%%%%%%%%%%%%%%%%%%%%%%%%%%%%%%%%%%%%%%%%%%%%%%%
%
\section*{Software}

\textcolor{black}{
The R programs developed for estimating rate matrices, testing accuracy of
the likelihood maximisation, significance testing, 
and calculating heterozygosity in Sections~\ref{sec:Application} and
\ref{sec:Simulation}
are available at \url{https://github.com/cjb105/RateMatrixEstimation}.
}
%
%%%%%%%%%%%%%%%%%%%%%%%%%%%%%%%%%%%%%%%%%%%%%%%%%%%%%%%%%%%%
%				Appendix
\appendix
\section{Appendix, Expectation-Maximisation Algorithm algorithm}\label{section:EM_algorithm}
\label{EMAlgorithm}

Here we provide the expectation-maximisation (EM) algorithm that can be used to estimate $e'$ and $b'$ from the conditional log-likelihood in Eq.~(\ref{eq:cond_ll_eb}).
Let the unknown auxiliary variable $z_{AC}(y)$ count the number of mutations from $A$ to $C$, the variable $z_{TG}(y)$ those from $T$ to $G$, and similarly for $z_{GA}(y)$ and $z_{CT}(y)$. As a logical consequence, $L_{AC}(y)+L_{TG}(y)-z_{AC}(y)-z_{TG}(y)$ counts the number of mutations from $C$ to $A$ and from $G$ to $T$. Then the conditional log-likelihood can be written as:
\begin{equation}
\begin{split}
\label{eq:cond_ll_em}
&\log(\Pr(L_{AC}(y),L_{TG}(y),L_{GA}(y),L_{TC}(y),z_{AC}(y),z_{TG}(y),z_{GA}(y),z_{TC}(y)\mid M, e',b')=\\
&\qquad const +\sum_{y=1}^{M-1} (z_{AC}(y)+z_{TG}(y)) \log\bigg(\frac{e'}{y}\bigg) \\ 
&\qquad+\sum_{y=1} ^{M-1}(L_{AC}(y)+L_{TG}(y)-z_{AC}(y)-z_{TG}(y))\log\bigg(\frac{1-b'}{M-y}\bigg)\\
&\qquad+\sum_{y=1}^{M-1} (z_{GA}(y)+z_{CT}(y)) \log\bigg(\frac{b'}{y}\bigg)\\ 
&\qquad+\sum_{y=1} ^{M-1}(L_{GA}(y)+L_{CT}(y)-z_{GA}(y)-z_{CT}(y))\log\bigg(\frac{1-e'}{M-y}\bigg)\,.
\end{split}
\end{equation}

The expectation step of the EM algorithm constitutes taking the expectation of Eq.(~\ref{eq:cond_ll_em}). 

It is helpful to look at the conditional expected values of the groups of auxiliary variables $z_{AC}(y)+z_{TG}(y)$, $L_{AC}(y)+L_{TG}(y)-z_{AC}(y)-z_{TG}(y)$, $z_{CA}(y)+z_{GT}(y)$, and $L_{GA}(y)+L_{CT}(y)-z_{GA}(y)-z_{CT}(y)$ separately:

The expectation of $z_{AC}(y)+z_{TG}(y)$ given $L_{AC}(y)+L_{TG}(y)$ corresponds to the mean of a binomial distribution with sample size $L_{AC}(y)+L_{TG}(y)$ and probability $p(y)=\frac{\frac{e'}{y}}{\frac{ e'}{y}+\frac{1-b'}{M-y}}$:
\begin{equation}
    \E(z_{AC}(y)+z_{TG}(y)\mid L_{AC}(y)+L_{TG}(y))=(L_{AC}(y)+L_{TG}(y))\frac{\frac{e'^t}{y}}{\frac{ e'^t}{y}+\frac{1-b'^t}{M-y}}\,.
\end{equation}
The other expectations follow analogously.

These can be plugged into Eq.~(\ref{eq:cond_ll_em}) to give:
\begin{equation}\label{eq:Q_of_e_b}
\begin{split}
&\operatorname{Q}(b',e'\mid b'^t,e'^t) =\\
&\qquad +\sum_{y=1}^{M-1} (L_{AC}(y)+L_{TG}(y)) \frac{\frac{e'^t}{y}}{\frac{e'^t}{y}+
\frac{1-b'^t}{M-y}} \log(e')\\
&\qquad +\sum_{y=1} ^{M-1}  (L_{AC}(M-y)+L_{TG}(M-y)) \frac{\frac{1-b'^t}{M-y}}{\frac{ e'^t}{y}
+\frac{1-b'^t}{M-y}}\log(1-b')\\
&\qquad +\sum_{y=1} ^{M-1} (L_{GA}(y)+L_{CT}(y))\frac{\frac{1-b'^t}{M-y}}{\frac{1-b'^t}{y}+
\frac{e'^t}{M-y}} \log(b') \\
&\qquad+\sum_{y=1}^{M-1}(L_{GA}(M-y)+L_{CT}(M-y)) \frac{\frac{e'^t}{M-y}}{\frac{1-b'^t}{y}
+\frac{e'^t}{M-y}}\log(1-e')\,.
\end{split}
\end{equation}

This leaves the maximisation step: To determine the overall iteration scheme for the parameter updates we solve the appropriate derivatives of $\operatorname{Q}(b',e'\mid b'^t,e'^t)$.
The overall iteration scheme is then given by:
\begin{equation}
    \begin{split}
        \hat e'^{t+1}&=\frac{\sum_{y=1}^{M-1} (L_{AC}(y)+L_{TG}(y)) \frac{\frac{e'^t}{y}}{\frac{e'^t}{y}+\frac{1-b'^t}{M-y}}}{\sum_{y=1}^{M-1} (L_{AC}(y)+L_{TG}(y)) \frac{\frac{e'^t}{M-y}}{\frac{e'^t}{y} +\frac{1-b'^t}{M-y}}+\sum_{y=1}^{M-1} (L_{GA}(M-y)+L_{CT}(M-y)) \frac{\frac{1-e'^t}{y}}{\frac{1-e'^t}{y}+\frac{b'^t}{M-y}}}\\
        \hat b'^{t+1}&=\frac{\sum_{y=1}^{M-1} (L_{GA}(y)+L_{CT}(y)) \frac{\frac{b'^t}{y}}{\frac{b'^t}{y}+\frac{1-e'^t}{M-y}}}{\sum_{y=1}^{M-1} (L_{GA}(y)+L_{CT}(y)) \frac{\frac{b'^t}{y}}{\frac{b'^t}{y}+\frac{1-e'^t}{M-y}}+\sum_{y=1}^{M-1} (L_{AC}(M-y)+L_{TG}(M-y)) \frac{\frac{1-b'^t}{M-i}}{\frac{e'^t}{y} +\frac{1-b'^t}{M-y}}}\,.
    \end{split}
\end{equation}
Cyclical calculation of estimators guarantees convergence towards a local maximum by properties of the EM algorithm. In this case, the local maximum is also the global maximum by the argument in the main text.

%
%%%%%%%%%%%%%%%%%%%%%%%%%%%%%%%%%%%%%%%%%%%%%%%%%%%%%%%%%%%%
%

\section*{Acknowledgments}

CV's research is supported by the Austrian Science Fund (FWF): DK W1225-B20; LCM's by the School of Biology at the University of St.Andrews. 

\section*{References}

\bibliographystyle{elsarticle-harv}\biboptions{authoryear}
%\bibliography{refs}  

%%%%%%%%%%%%%%%%%%%%%%%%%%%%%%%%%%%%%%%%%%%%%%%%%%%%%%%%%%%%%%%%%%%%%%%%%%%%%%%%

\end{document}